\documentclass[preprintnumbers, superscriptaddress, floatfix, letterpaper, onecolumn,aps,prd,epsfig,nofootinbib,natbib,longbibliography]{revtex4-2}
\usepackage[colorlinks,linkcolor=blue,citecolor=magenta,urlcolor=blue]{hyperref}
\usepackage{bm,graphicx,dcolumn,epstopdf,epsf, latexsym,mathbbol, amssymb,amsmath,color,slashed, mathrsfs,mathcomp,simplewick}
\usepackage{graphicx}% Include figure files
\usepackage{tabularx}
\usepackage{dcolumn}% Align table columns on decimal point
\usepackage{bm}% bold math
\usepackage{orcidlink} % For \orcid command

\usepackage{graphicx}
\usepackage{hyperref}

\usepackage{booktabs}  % in preamble for tables

\newcommand{\orcid}[1]{%
  \href{https://orcid.org/#1}{%
    \includegraphics[width=10pt]{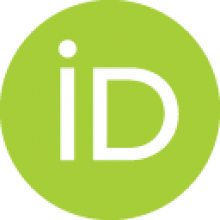}%
  }%
}

\usepackage{array}
\usepackage{booktabs}
\usepackage{tabu}
\usepackage{dcolumn}

\usepackage{amsmath}
\usepackage{amsfonts}
\usepackage{amssymb}
\usepackage{graphicx}
\usepackage{subfigure}
\usepackage{graphicx}% Include figure files
\usepackage{bm}% bold math
\usepackage{xcolor}

\graphicspath{{figs/}} % Specify the path to the figures folder

\begin{document}

\title{
% Thresholds and Abundances of Primordial Black Holes in Non-Standard Thermal History
Primordial Black Hole Abundances and Scalar Induced Gravitational Waves from Finite-Width Power Spectra in a Stiff Thermal History %Cosmologies
}
\author{Abolhassan Mohammadi\orcid{0000-0003-1228-9107}}%
    \email{abolhassanm@hnit.edu.cn}
    \affiliation{%
School of Science, Hunan Institute of Technology, Hengyang 421002, China.
}%

\author{Yogesh\orcid{0000-0002-7638-3082}}
    \email{yogeshjjmi@gmail.com}
    \affiliation{%
Centre for Cosmology and Science Popularization (CCSP), SGT University, Gurugram, Haryana 122505, India.
}%

\author{Cristian Joana\orcid{0000-0003-4642-3028}}
\email{cristian.joana@ucas.ac.cn}
\affiliation{International Centre for Theoretical Physics Asia-Pacific,
University of Chinese Academy of Sciences, 100190 Beijing, China}
\affiliation{Center for Theoretical Physics, School of Physics and Optoelectronic Engineering,\\ Hainan University, Haikou 570228, China}

\author{Hongwei Tan}%
    \email{honweitan@hnit.edu.cn}
    \affiliation{%
School of Science, Hunan Institute of Technology, Hengyang 421002, China.
}%

\author{Ying-li Zhang\orcid{0000-0003-0396-1408}}
\email{yingli@tongji.edu.cn}
\affiliation{School of Physics Science and Engineering, Tongji University, Shanghai 200092, China}
\affiliation{Institute for Advanced Study of Tongji University, Shanghai 200092, China}
\affiliation{Asia Pacific Center for Theoretical Physics, Pohang 37673, Korea}
\affiliation{Kavli Institute for the Physics and Mathematics of the Universe (WPI), The University of Tokyo Institutes for Advanced Study, The University of Tokyo, Chiba 277-8583, Japan}
\affiliation{Center for Gravitation and Cosmology, Yangzhou University, Yangzhou 225009, China}

% \author{Misao Sasaki}
% \email{misao.sasaki@apctp.org}
% \affiliation{Asia Pacific Center for Theoretical Physics (APCTP), Pohang 37673, Republic of Korea}
% \affiliation{Kavli Institute for the Physics and Mathematics of the Universe (WPI),
% UTIAS The University of Tokyo, Kashiwa, Chiba 277-8583, Japan.}
% \affiliation{Center for Gravitational Physics and Quantum Information, Yukawa Institute for Theoretical Physics, Kyoto University, Kyoto 606-8502, Japan}

%---------------------
\begin{abstract}
We study the formation of primordial black holes (PBHs) from large primordial perturbations that re-enter the horizon during an epoch with equation of state $\mathrm{w}\geq1/3$. We consider a log-normal curvature power spectrum of finite width $\Delta$ and determine the collapse amplitude by numerical-relativity simulations of a self-gravitating perfect fluid. Threshold scans are performed for five values of $\mathrm{w}$ and five spectral widths, and the resulting numerical thresholds are used in the PBH abundance and scalar-induced gravitational-wave (SIGW) calculations.
{ For comparison, we also evaluate two calibrations of the semi-analytical \(\mathrm{w}q\)-prescription and find that, while they may reproduce the numerical threshold within a \(2.8\text{–}8\%\) error, this precision is often insufficient for PBH abundance computations.}
% They reproduce the numerical threshold within a $2.8$-–$8\%$ error, this precission is often insuficient for PBH abundances computations.} 
%It reproduces the numerical trend close to radiation domination and for nearly monochromatic profiles, but it is not a reliable threshold estimator for generic finite-width profiles in stiffer backgrounds 
We show how collapse thresholds increase with both $\Delta$ and $\mathrm{w}$, changing the curvature amplitude required for PBHs to constitute all of the dark matter and, consequently, the normalization of the associated SIGW signal.

\end{abstract}

\maketitle

%++++++++++++++++++++++++++++++++++++
%++++++++++++++++++++++++++++++++++++
%++++++++++++++++++++++++++++++++++++
%+++++ 1. Introduction
%++++++++++++++++++++++++++++++++++++
%++++++++++++++++++++++++++++++++++++
%++++++++++++++++++++++++++++++++++++
\section{Introduction}
\label{sec:intro}

Primordial black holes (PBHs) were first proposed by Zel'dovich and Novikov in 1967~\cite{Zeldovich:1967lct}. Later on, the theory was worked out more formally by Hawking and Carr~\cite{Hawking:1971ei,Carr:1974nx,Carr:1975qj}, who showed that the resulting PBH mass is roughly equal to the horizon mass at the moment of collapse~\cite{Carr:1974nx,Carr:1993aq}. Due to their non-stellar origin, PBHs are not bound by the mass limit imposed by the Chandrasekhar limit~\cite{Chandrasekhar:1931ih}. Theoretically, they can be of any mass from a few grams to thousands of solar masses. Interest in PBHs picked up sharply after LIGO and Virgo detected GW signals from binary (P)BHs mergers~\cite{LIGOScientific:2016aoc,LIGOScientific:2016dsl,LIGOScientific:2016sjg,LIGOScientific:2016wyt,LIGOScientific:2017bnn,LIGOScientific:2017vox,LIGOScientific:2017ycc}. Several of the observed events involve masses that are hard to produce through standard stellar evolution, and a primordial origin for at least some of these black holes has been proposed~\cite{Bird:2016dcv,Clesse:2016vqa,Sasaki:2016jop,Fernandez:2019kyb,DeLuca:2020qqa}. More recently, PTA experiments, NANOGrav, EPTA, PPTA, and CPTA reported a common-spectrum stochastic GW background in the nanohertz band~\cite{NANOGrav:2023gor,EPTA:2023fyk,2013PASA...30...17M,Hobbs:2013aka,Xu:2023wog}. A compelling interpretation of this signal is the SIGW background produced by large primordial curvature perturbations~\cite{Kohri:2018awv,Espinosa:2018eve,Domenech:2021ztg,Choudhury:2024aji,Maiti:2026hsn}. The fact that PBHs, PTA signals, and dark matter (DM) can all point to the same set of enhanced small-scale perturbations motivates studying them in a single framework.

PBHs with mass below $\sim 10^{15}~\rm g$ are expected to have evaporated by today through Hawking radiation~\cite{Hawking:1971ei}, while heavier ones persist and therefore contribute to the DM density. In the mass window $10^{-16} M_\odot \lesssim M_{\rm PBH} \lesssim 5 \times 10^{-12} M_\odot$, current observational constraints still allow PBHs to account for all of DM~\cite{Carr:2020gox,Carr:2021bzv,Escriva:2022duf}. The lighter end of the spectrum is constrained by galactic and extragalactic $\gamma$-ray observations, while the heavier end is bounded by microlensing surveys and GW data~\cite{Carr:2021bzv,Laha:2019ssq}.

Several mechanisms for PBH production have been studied, including bubble nucleation at first-order phase transitions~\cite{Crawford:1982yz,Hawking:1982ga}, bubble collapse~\citep{Garriga:2015fdk,Deng:2017uwc,Wang:2025hwc,Joana:2026myf}, underdense/void and shell collapse~\citep{Joana:2026myf,Joana:2025gqf,Germani:2025hcu}, collapse of cosmic strings~\cite{HAWKING1989237,Blanco-Pillado:2017rnf} or domain walls~\cite{Rubin:2001yw,Garriga:2015fdk}, and formation in modified gravity theories~\cite{Kawai:2021edk,Papanikolaou:2022hkg,Banerjee:2022xft,Gangopadhyay:2026mck,Yogesh:2025hll,Ashrafzadeh:2024oll,Solbi:2024zhl,Ashrafzadeh:2023ndt,HosseiniMansoori:2023mqh}. For the PBHs formation, we focus here on the collapse of large curvature perturbations generated during inflation~\citep{Carr:1993aq,Kawasaki:1997ju,Yokoyama:1998pt,Germani:2017bcs,Ballesteros:2017fsr,Pi:2017gih,Biagetti:2018pjj,Atal:2019erb,Solbi:2021wbo,Pi:2022zxs,Wang:2024vfv,Inui:2024fgk,Cacciapaglia:2025xqd,Wang:2025lti,Yuwen:2026hxu,Gangopadhyay:2026xqj,Braglia:2020eai,Mohammadi:2025avz,Maiti:2025ijr}. During inflation, scalar perturbations are generated, stretched out, cross the Hubble horizon, and freeze as super-horizon fluctuations. After inflation ends, the comoving Hubble radius grows, and these modes re-enter the horizon. On re-entering, if the perturbations are large enough, they collapse and form a PBH, with mass determined by the horizon mass at re-entry~\cite{Young:2019yug,PhysRevD.50.7173,Garcia-Bellido:2017mdw}.

The challenge is that this requires the primordial power spectrum to be enhanced by roughly seven orders of magnitude on sub-CMB scales, reaching $\mathcal{O}(10^{-2})$ compared to the observed CMB amplitude $\mathcal{P}_\zeta(k_\star) \simeq 2.1 \times 10^{-9}$~\cite{Planck:2018jri,Sato-Polito:2019hws}. In canonical single-field inflation, producing such an enhancement requires a transient ultra-slow-roll (USR) phase, where the scalar field traverses a near-flat region of the potential and the first Hubble slow-roll parameter $\epsilon_1$ drops by many orders of magnitude~\cite{Garcia-Bellido:2017mdw,Kinney:2005vj,Ragavendra:2020sop,Biagetti:2018pjj,Kim:2025dyi}. Other approaches introduce localised potential features such as bumps or steps~\cite{Hazra:2010ve,Mishra:2019pzq}. The USR (or bump) must be relatively brief so that it does not disturb the large-scale spectrum, which is tightly constrained by CMB observations.
Here $H=\dot a/a$ is the Hubble parameter, $a(t)$ is the scale factor, and a dot denotes a derivative with respect to cosmic time $t$.

Whether a given overdense region actually collapses into a PBH is determined by a threshold, and finding the right threshold value is critically important.
The abundance depends on it exponentially, so even a small error feeds through into a large change in the predicted number of PBHs. The first and simplest estimate of the threshold goes back to Carr~\cite{Carr:1975qj}, who used a Jeans-length argument to set the critical density contrast at $\delta_c \simeq 1/3$ in a radiation background, the minimum overdensity needed to overcome the pressure gradient. However, such a single value carries no information about the shape of the collapsing region.

A more reliable measure is the compaction function~\cite{Shibata:1999zs}, which gives the excess mass within a given areal radius and keeps the dependence on the perturbation profile. Numerical relativity (NR) simulations have since confirmed this, showing that the threshold obtained from the compaction function changes considerably with the perturbation profile~\cite{Musco:2018rwt,Escriva:2019phb}.

For the analysis of PBH formation, we use the peak theory~\cite{Bardeen:1985tr,Peacock:1990zz} rather than the Press-Schechter formalism~\cite{1974ApJ...187..425P}. The Press-Schechter method gives a quick statistical estimate but treats density fluctuations only through their variance, losing information about the spatial profile of the collapsing region. Peak theory works with the number density of local maxima of the curvature perturbation field, treated as a Gaussian random field, and gives a direct connection between the power spectrum and the shape of the collapsing profile~\cite{Yoo:2018kvb,Germani:2019zez,Young:2020xmk}. In this framework, the collapse threshold $\delta_c$ depends on the profile shape through the compaction function $\mathcal{C}(r)$~\cite{Shibata:1999zs,Escriva:2019phb}, which measures the mass excess inside a sphere of areal radius $R(r)$. 
We therefore determine the collapse thresholds across the full range of spectral widths and equations of state considered here using full NR simulations.

A common simplification is to take the curvature power spectrum as narrow, often nearly monochromatic. This reduces the problem to a single scale that determines the size of every collapsing region, simplifying the calculation. However, such a sharp spectrum is hard to produce in a realistic inflationary model, and the resulting abundances may not carry over to physical cases~\cite{Germani:2023ojx}. A spectrum of finite width is the generic situation. Modes across a band of scales re-enter the horizon and collapse, so PBHs are produced over a range of masses, and the resulting abundance depends on the width of the spectrum. At fixed variance, a broader spectrum gives a lower abundance, since spreading the power over more scales lowers the peak amplitude~\cite{Pi:2024jwt}. Treating this correctly requires the full-scale dependence of the spectrum, together with a window function that selects the scales relevant to each PBH mass~\cite{Pi:2024jwt}.

Another extension of this work concerns the thermal history between the end of inflation and the start of Big Bang nucleosynthesis (BBN), which involves uncertainties in the effective equation of state, $\mathrm{w}$, of the universe. The reheating temperature can range over many orders of magnitude from the end of inflation to the initiation of BBN~\cite{Bhattacharya:2023ztw}. The universe may pass through a phase dominated by some species $\Phi$ with $\rho_\Phi \propto a^{-3(1+\mathrm{w})}$ where $\mathrm{w} \neq 1/3$ between the end of reheating and start of BBN. An early matter-dominated phase ($\mathrm{w} \approx 0$) arises naturally from moduli fields in string-motivated cosmologies~\cite{Allahverdi:2020bys,Coughlan:1983ci,Starobinsky:1994bd}, while kinetic energy domination ($\mathrm{w} \approx 1$) is generic in quintessential inflation scenarios~\cite{Peebles:1998qn,Ahmad:2019jbm}. BBN constrains the radiation epoch to begin $T \geq 4~\rm MeV$~\cite{Bhattacharya:2023ztw,Kawasaki:1999na,Hasegawa:2019jsa}. At the time of the horizon re-entry, a phase with $\mathrm{w} \neq 1/3$ affects the mass-scale relation, collapse threshold, and the final abundance~\cite{Escriva:2022pnz,Bhattacharya:2019bvk,Bhattacharya:2020lhc,Bhattacharya:2023ztw,Yogesh:2025hll}. Restricting to $\mathrm{w} = 1/3$ may miss a large part of the physically interesting parameter space.

In this work, we study the formation of PBHs from a wide curvature power spectrum, which we take to have a log-normal shape, and we allow the collapse to occur in a general epoch with an equation of state $\mathrm{w} \geq 1/3$ rather than fixing the background to radiation.  We treat the width of the power spectrum, $\Delta$, and the equation of state, $\mathrm{w}$, as free parameters and study how they affect the thresholds, the PBH mass function, and the total abundance. 
Collapse thresholds are computed using NR simulations and compared to the semi-analytical estimate from the $q$-function method through the $(\Delta,\mathrm{w})$ parameter space. Furthermore, we compute the SIGWs generated by these perturbations and show how the signal depends on $\Delta$ and $\mathrm{w}$.

The paper is organized as follows. In Section~\ref{sec:Ps}, we introduce the log-normal power spectrum. Section~\ref{sec:peak} reviews peak theory and compares the semi-analytical and NR thresholds. Section~\ref{sec:abundance} presents the PBH mass function and abundance, and Section~\ref{sec:sigw} gives the corresponding SIGW signals. Section~\ref{sec:conclusion} contains our conclusions. Technical details on the NR simulations, including formulation, initial data, horizon finder, and constraint checks, are documented in Appendix~\ref{app:NR}.

%----------------------------
%----------------------------
%----    The Curvature Power Spectrum
%----------------------------
%----------------------------
\section{The Curvature Power Spectrum}
\label{sec:Ps}

In the study of PBHs, all results are determined by the statistics of the comoving curvature perturbation $\zeta$. For Gaussian fluctuations, this is fully described by the dimensionless power spectrum $\mathcal{P}_\zeta(k)$. One way to obtain this power spectrum is to choose an inflationary model and solve the mode equation for $\zeta$ over a wide range of the mode scale $k$. Here, we take a different, more direct path, which is also a natural choice for the study of PBHs. Instead of fixing a specific model, we use an analytic form for the power spectrum that captures the main features that any model of PBHs should have, i.e., a nearly scale-invariant part on the large scales seen by the CMB, and an enhancement with finite width on much smaller scales, where the collapse takes place. The power spectrum is then taken as
\begin{equation}\label{Ps_lognormal}
    \mathcal{P}_\zeta(k) = A_s\left(\frac{k}{k_\star}\right)^{n_s-1}
    + \frac{A}{\sqrt{2\pi}\,\Delta}\,
    \exp\!\left[-\frac{\ln^2(k/k_p)}{2\Delta^2}\right] ~. 
\end{equation}
The first term is the usual near-scale-invariant component tied to the CMB, with amplitude $A_s \simeq 2.1\times10^{-9}$ and spectral index $n_s$ at the pivot scale $k_\star = 0.05~\mathrm{Mpc}^{-1}$~\cite{Planck:2018jri,ACT:2025fju}. The second term is a log-normal bump, the form most often used to represent a finite-width peak in studies of PBH formation and the associated induced gravitational waves~\cite{Pi:2020otn}. It is specified by three quantities, namely the amplitude $A$, the comoving scale $k_p$ at which the enhancement is centered, and the dimensionless width $\Delta$.

The log-normal piece is a Gaussian in $\ln k$ centered on $k=k_p$. Its normalization is chosen so that its area in logarithmic wavenumber equals $A$,
\begin{equation}\label{Avariance}
    \int \frac{\mathrm{d}k}{k}\,
    \frac{A}{\sqrt{2\pi}\,\Delta}\,
    \exp\!\left[-\frac{\ln^2(k/k_p)}{2\Delta^2}\right] = A \;.
\end{equation}
So $A$ sets the total power in the enhanced region, while the width $\Delta$ sets how that power is spread in scale. At fixed $A$, the peak height goes as $A/(\sqrt{2\pi}\,\Delta)$, so a wider spectrum is also a lower one. In the limit $\Delta\to0$ the bump becomes a Dirac delta, and we recover the monochromatic spectrum used in much of the earlier work. The peak sits at $k_p \gg k_\star$, so it is well separated from the large-scale plateau and does not affect the CMB bounds.

\begin{figure}
    \centering
    \includegraphics[width=0.47\linewidth]{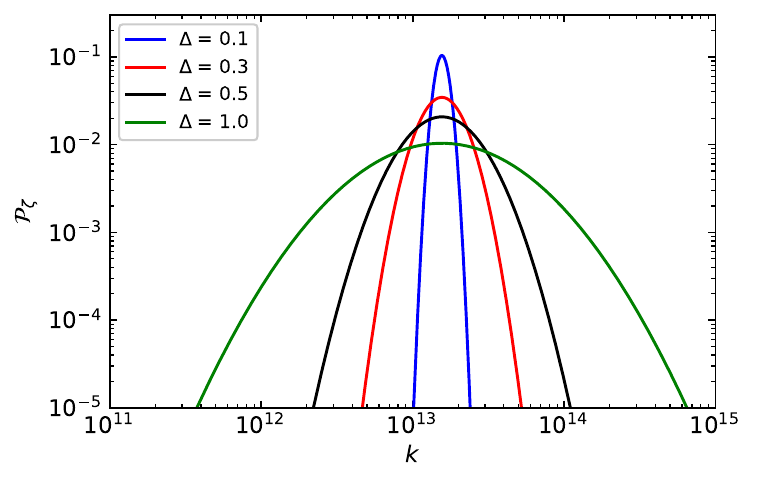}
    \caption{The curvature power spectrum of Eq.~\eqref{Ps_lognormal} for four widths, $\Delta=0.1$, $0.3$,  $0.5$ and $1$ at fixed amplitude $A = 2.6 \times 10^{-2}$ and peak scale $k_p = 1.56\times10^{13}\,{\rm Mpc}^{-1}$. The near-scale-invariant component dominates at large scales and keeps the spectrum consistent with the CMB, while the log-normal bump governs the small-scale power that drives PBH formation. With the amplitude $A$ held fixed, a wider power spectrum leads to a smaller peak amplitude, since the same power is distributed over a wider range of wavenumbers.}
    \label{fig:Ps}
\end{figure}

The log-normal form is more than just a convenient case. It is a reasonable approximation to the enhanced power spectrum produced by single-field models. In these models of inflation, a short non-attractor (ultra-slow-roll) stage between two slow-roll phases amplifies the small-scale power, and the resulting peak has a finite width set by how long that stage lasts~\cite{Garcia-Bellido:2017mdw,Biagetti:2018pjj}. Such spectra are often written as a broken power law, and a log-normal of width $\Delta$ provides a simple description of the same peak~\cite{Byrnes:2018txb,Pi:2020otn}. Since the width is not fixed by a single model, we keep $\Delta$ free and study a few representative values, from a nearly monochromatic peak ($\Delta=0.1$) to broad ones ($\Delta=0.5$ and $\Delta=1.0$), so the effect of the width can be seen directly. Fig.~\ref{fig:Ps} displays a schematic of the power spectrum for different values of the width parameter $\Delta$ and for fixed values of $A$ and $k_p$. It is realized that, for a fixed value of $A$, widening the power spectrum results in a lower peak because the same power is now distributed over a wider range of $k$ modes. 
%-----------------------
%-----------------------
The spectrum in Eq.~\eqref{Ps_lognormal} is the only input required for the rest of the analysis. The variance, the higher spectral moments, and the correlation functions used in peak theory are all integrals of $\mathcal{P}_\zeta(k)$ over a suitable window, as shown in the next section.

%%%%%%%%%%%%%%%%%%%%%%%%%%%%%%%%%%%%%%%%%%%%%%%%%%%%%%%%%%%%%
%%%%%%%%%%%%%%     3. Profile in peak theory     %%%%%%%%%%%%%%%%%%%%%%
%%%%%%%%%%%%%%%%%%%%%%%%%%%%%%%%%%%%%%%%%%%%%%%%%%%%%%%%%%%%%
%%%%%%%%%%%%%%%%%%%%%%%%%%%%%%%%%%%%%%%%%%%%%%%%%%%%%%%%%%%%%
\section{Peak Theory and PBH Formation Threshold}
\label{sec:peak}
To compute the abundance of PBHs, a statistical description of how inflationary curvature perturbations collapse upon horizon re-entry is needed. The Press-Schechter approach is the most widely used method for estimating the PBH fraction, which involves integrating the tail of the density distribution over a certain threshold~\citep{1974ApJ...187..425P}. While this method is straightforward to use, it has its own flaws. The Press-Schechter discards all information regarding the collapsing region's spatial profile, treating peaks with the same amplitude as statistically identical, regardless of shape. On the other hand, peak theory~\citep{Bardeen:1985tr,Peacock:1990zz} 
estimates the abundance from the number density of local maxima of a Gaussian random field and retains statistical information about their curvature and spatial profiles~\citep{Yoo:2018kvb,Germani:2019zez,Young:2020xmk}.
This allows the abundance calculation to account for the dependence of PBH formation on the peak's shape.  
The corresponding collapse threshold is obtained either from NR simulations or from a calibrated analytical prescription.
% presents a more detailed and robust approach by accounting for the number density of local maxima of a Gaussian random field and preserving the shape of the profile at each peak, enabling a more reliable estimation of the collapse threshold~\citep{Yoo:2018kvb,Germani:2019zez,Young:2020xmk}. Numerical simulations performed in Ref.~\citep{Escriva:2022pnz} have shown excellent agreement with peak theory, with discrepancies remaining at only the few-percent level. 

%++++++++++++++++++++++++++++++
%++++++++++++++++++++++++++++++
%++++++++++++++++++++++++++++++
\subsection{Curvature perturbation profile in peak theory}
\label{sec:peak_theory}

The spatial component of the perturbed FLRW metric on the comoving slice reads
\begin{equation}
    ds_3^2 = a^2(t)\,e^{2\zeta(r)}\!\left(dr^2 + r^2\,d\Omega^2\right),
    \label{metric3}
\end{equation}
where $\zeta(r)$ denotes the comoving curvature perturbation, $r$ represents the radial comoving coordinate, and $d\Omega^2 = d\theta^2 + \sin^2\!\theta\,d\varphi^2$. We restrict ourselves to the spherically symmetric scenario, which is a suitable approximation because the rare, high peaks responsible for PBH production are almost spherical~\citep{Bardeen:1985tr}\footnote{For the formation of non-spherical PBHs, readers may go through the Refs.\citep{Escriva:2024lmm, Escriva:2024aeo}}. The field $\zeta$ is treated as a Gaussian random variable with a distribution
\begin{equation}
    \mathbb{P}_G(\zeta) = \frac{1}{\sqrt{2\pi}\,\sigma_\zeta}\exp\!\left(-\frac{\zeta^2}{2\sigma_\zeta^2}\right),
    \label{PDF}
\end{equation}
where  $\sigma_\zeta^2$ signifies the variance, which is computed by the power spectrum,
\begin{equation}
    \sigma_\zeta^2 = \int\frac{\mathrm{d}k}{k}\,\mathcal{P}_\zeta(k)\,\widetilde{W}^2(k,k_p),\qquad \widetilde{W}(k,k_p) = \exp\!\left[-\frac{1}{2}\!\left(\frac{k}{k_p}\right)^{\!2}\right],
    \label{sigma_zeta}
\end{equation}
with $k_p$ being the scale where the power spectrum peaks and $\widetilde{W}(k,k_p)$ a Gaussian window adopted for smoothing of the perturbations~\citep{Pi:2024ert}.

Following the formalism developed in Refs.~\citep{Yoo:2020dkz,Kitajima:2021fpq,Pi:2024ert}, we construct the peak statistics using $\nabla^2\zeta$ as the underlying Gaussian random field. This choice is motivated by two considerations. At linear order, $-\nabla^2\zeta \propto \delta\rho/\rho$, so its local maxima correspond to overdense regions. At the same time, because PBH formation is governed by local physics, the long-wavelength part of $\zeta$ can be absorbed into a redefinition of the local scale factor and plays no role in the collapse dynamics~\citep{Yoo:2020dkz}.

Each peak is characterized by two parameters, the height $\mu \equiv -\nabla^2\zeta|_{r=0} \cdot \sigma_1^2/ \sigma_2^2$ and the dimensionless curvature $K^2 \equiv -\mu^{-1}\nabla^2(-\nabla^2\zeta)|_{r=0}\cdot\sigma_2/\sigma_4$. The spectral moments computed from the power spectrum are defined as
\begin{equation}
    \sigma_n^2 = \int\frac{ \mathrm{d}k}{k}\,k^{2n}\,\mathcal{P}_\zeta(k)\,\widetilde{W}^2(k,k_p),
    \label{spectral_moments}
\end{equation}
also, the two-point correlation functions are
\begin{equation}
    \psi_n(r) = \frac{1}{\sigma_n^2}\int\frac{\mathrm{d}k}{k}\,k^{2n}\frac{\sin(kr)}{kr}\,\mathcal{P}_\zeta(k)\,\widetilde{W}^2(k,k_p),
    \label{twopoint}
\end{equation}
with auxiliary shape parameters $\gamma_n = \sigma_n^2/(\sigma_{n-1}\sigma_{n+1})$ and $R_n = \sqrt{3}\,\sigma_n/\sigma_{n+1}$ (defined for odd $n$)~\citep{Inui:2024fgk,Pi:2024ert}. Following ~\citep{Pi:2024ert} the peak profile ($\hat\zeta(r)$) can be defined as
%:\footnote{For detailed discussion, readers are suggested to go through Refs.~\citep{Yoo:2018kvb,Yoo:2020dkz}}
\begin{equation}
    \hat\zeta(r) = \frac{\mu}{1-\gamma_3^2}\!\left[\psi_1(r) + \frac{R_3^2}{3}\nabla^2\psi_1(r) - \frac{K^2}{\gamma_3}\!\left(\gamma_3^2\psi_1(r) + \frac{R_3^2}{3}\nabla^2\psi_1(r)\right)\right]+\zeta_\infty,
    \label{zeta_profile}
\end{equation}
where the integration constant $\zeta_\infty$ can be set to zero by absorbing it into the redefinition of the background scale factor. For simplicity, we fix $K=\sqrt{\gamma_3}$, corresponding to the most probable value of the peak-curvature parameter, so that $\mu$ is the only free parameter characterizing the profile. We, however, note that although deviations from this value are Gaussian-suppressed in the peak distribution~\citep{Pi:2024ert}, the dependence of the collapse threshold on $K$ can nevertheless make peaks associated with less probable profiles relevant.

%++++++++++++++++++++++++++++
%++++++++++++++++++++++++++++
%++++++++++++++++++++++++++++
\subsection{Compaction function and semi-analytical thresholds}
\label{sec:compaction}

One of the most fundamental steps in determining whether a peak can collapse into a black hole is setting the threshold. The key quantity used to compute the threshold value is the compaction function. It measures the mass excess of the collapsing region over a given areal radius, and is defined as 
\citep{Shibata:1999zs,Escriva:2019phb,Harada:2015yda,Yoo:2018kvb,Musco:2020jjb},
\begin{equation}
    \mathcal{C}(r) = 2\,\frac{M(r,t)-M_b(r,t)}{R(r,t)},
    \label{compaction_def}
\end{equation}
where $M(r,t)$ is the enclosed Misner--Sharp mass and $R(r,t)=a(t)r e^{\zeta(r)}$ is the areal radius. The background FLRW mass is $M_b = \frac{4\pi}{3}\rho_b R^3$, with $\rho_b$ the background energy density. Equation~\eqref{compaction_def} uses geometrized units $G=c=1$. On the comoving slices and by integrating at the leading order of the gradient expansion, the compaction function can be obtained in terms of the curvature perturbations 

\begin{figure}[h]
    \centering
    \includegraphics[width=0.80\linewidth]{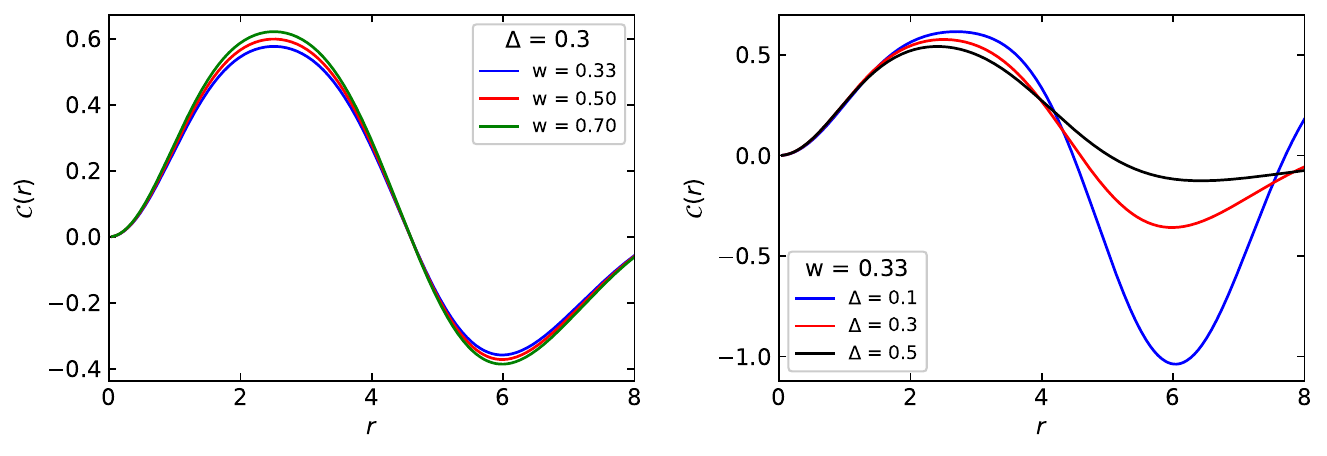}
    % \caption{The figure shows the behavior of the compaction function $\mathcal{C}(r)$ with the radial coordinate $r$ for three different equations of state ($\mathrm{w}$).}
    \caption{Compaction function $\mathcal{C}(r)$ for curvature profiles with amplitude $\mu=0.7$. The left panel fixes $\Delta=0.3$ and compares $w=1/3$, $1/2$, and $0.7$. The right panel fixes $w=1/3$ and compares $\Delta=0.1$, $0.3$, and $0.5$. 
    % Increasing $w$ raises the maximum compaction, while increasing $\Delta$ lowers and broadens the peak.
    }
    \label{fig:comp}
\end{figure}

\citep{Shibata:1999zs,Escriva:2019phb}
\begin{equation}\label{compaction_expr}
    \mathcal{C}(r) = f(\mathrm{w})\!\left(1-\!\left[1+r\zeta'(r)\right]^2\right),\qquad f(\mathrm{w}) \equiv \frac{3(1+ \mathrm{w})}{5+3 \mathrm{w}},
    % \mathcal{C}(r) = f(w)\!\left(1-\!\left[1+r\zeta'(r)\right]^2\right),\qquad f(w) \equiv \frac{3(1+w)}{5+3w}~.
\end{equation}
Here, the primes denote radial derivatives. We assume that the universe is dominated by a perfect fluid with energy density $\rho$ and pressure $P$ with a linear equation of state $\mathrm{w} = P/\rho$. As long as the equation of state parameter is taken as a constant, the above compaction function is time-independent, and it remains a function of $r$. The location of the peak of the compaction function is determined by $r_m$, and it can be obtained by solving the equation $\zeta'(r_m)+r_m\zeta''(r_m)=0$. Fig.\ref{fig:comp} displays the compaction function for different values of $\mathrm{w}$ (left panel) and $\Delta$ (right panel). One can see that, by increasing $\mathrm{w}$, the peak of the compaction function increases, as shown in the left panel. On the other hand, for the fixed value of $\mathrm{w}$, we have a lower peak as $\Delta$ increases; however, we have a flatter peak for a lower value of $\Delta$, as shown in the right panel of Fig.\ref{fig:comp}.
%--------------------------
%--------------------------
We restrict the abundance calculation to type-I fluctuations, for which $R'(r)\propto1+r\zeta'(r)>0$ and the areal radius is monotonic~\citep{Escriva:2019phb}. In contrast, type-II profiles contain a region with $1+r\zeta'(r)<0$~\citep{Harada:2024jxl,Uehara:2025idq,Escriva:2025rja,Shimada:2024eec}. For the curvature profiles considered here, the transition to type II occurs only at amplitudes well above the collapse threshold. These configurations therefore lie deeper in the suppressed tail of the amplitude distribution, making their contribution to the PBH abundance subdominant.

Numerical simulations performed in ~\citep{Escriva:2019phb} established a threshold criterion for PBH formation during a radiation-dominated epoch. They determined that a PBH can be formed as the volume average of the compaction function, 
\begin{equation}\label{Cavg}
    \bar{\mathcal{C}}_m = \frac{3}{R_m^3}\int_0^{R_m}\mathcal{C}(r)\,R^2(r)\,\mathrm{d}R ,
\end{equation}
satisfies the condition $\bar{\mathcal{C}}_m > \bar{\mathcal{C}}_{th} = 2/5$, where $R_m\equiv R(r_m)$ is the areal radius at the compaction maximum. This universal threshold only applies to the radiation phase~\citep{Escriva:2019phb}.

%++++++++++++++++++++++++++++
%++++++++++++++++++++++++++++
%++++++++++++++++++++++++++++
% \subsection{The $q$-Function Method and the Threshold for General $\omega$}
% \label{sec:qfunction}

Analogously, the $q$-function approach is another method for determining the threshold. The parameter $q$ measures the curvature of the compaction function at its maximum, which captures the dependence of the threshold on the shape of the profile. This parameter is defined as % ~\citep{Escriva:2019phb}
\begin{equation}\label{qfactor_tilde}
    q = -\frac{\mathcal{C}''(r_m)\, r_m^2}{4\,\mathcal{C}(r_m)\!\left[1-\mathcal{C}(r_m)/f(\mathrm{w})\right]}~,
\end{equation}
where $\mathcal{C}''(r_m) = d^2\mathcal{C}/dr^2|_{r_m}$. Note that the dependence on $\mathrm{w}$ cancels, leaving $q$ independent of $\mathrm{w}$.

A sharp peak with large pressure gradients has a large $q$, while a broad, flat peak has a small $q$. In radiation domination, $\mathrm{w}=1/3$, the threshold on the peak compaction can be approximated as a function of $q$ alone~\citep{Escriva:2019phb}.  We call this the (radiation-only) $q$-prescription,
\begin{equation}
    % \text{(radiation }q\text{-prescription)}\qquad
    \delta_c^{\rm rad}(q) = \frac{4}{15}\,e^{-1/q}\,\frac{q^{1-\tilde q}}{\Gamma(\tilde q)-\Gamma(\tilde q,1/q)},\qquad \tilde q \equiv \frac{5}{2q},
    \label{dc_rad}
\end{equation}
where $\Gamma(x)$ and $\Gamma(x,z)$ are the Euler Gamma function and the upper incomplete Gamma function, respectively. For a general equation of state with $\mathrm{w}\geq1/3$, the averaging region of the compaction function also depends on $\mathrm{w}$~\citep{Escriva:2020tak}. This gives a threshold depending on both $\mathrm{w}$ and the profile-shape parameter $q$, which we call the $\mathrm{w}q$-prescription,
{
\begin{equation}
    \text{(}\mathrm{w}q\text{-prescription)}\qquad
    \delta_c(\mathrm{w},q) = \frac{\bar{\mathcal{C}}_c(\mathrm{w})}{p(\mathrm{w},q)}\cdot\frac{1}{\left[(1-\alpha)^{3-2q}\,F_2(q,\mathrm{w})-F_1(q)\right]},
    \label{dc_gen}
\end{equation}
}
The functions $\bar{\mathcal{C}}_c(\mathrm{w})$ and $\alpha(\mathrm{w})$ specify the critical averaged compaction and the extent of the averaging region. Ref.~\citep{Escriva:2020tak} provides a calibration fitted to simulations over $0.1\leq q\leq30$. We refer to Eqs.~(28)--(29) of that reference as the $\mathrm{w}q$-fit prescription, intended for this low-$q$ range,
{
\begin{subequations}\label{wq_fit_calibration}
\begin{align}
    \text{(}\mathrm{w}q\text{-fit)}\qquad
    \bar{\mathcal{C}}_c^{\rm fit}(\mathrm{w})
    &= -0.140381 + 0.79538\,\arctan(1.23593\,\mathrm{w}^{0.357491}),\label{Cc_fit_numeric}\\[4pt]
    \alpha_{\rm fit}(\mathrm{w})
    &= 2.00804 - 1.10936\,\arctan(10.2801\,\mathrm{w}^{1.113}).\label{alpha_fit_numeric}
\end{align}
\end{subequations}
}
To cover the full range of positive $q$, the same reference constructs an alternative calibration using the $q\to0$ threshold to determine $\bar{\mathcal{C}}_c$ and the $q\to\infty$ condition $\delta_c\to f(\mathrm{w})$ to determine $\alpha$. This calibration is described by Eqs.~(37)--(38) of that same reference, and we refer to them as the $\mathrm{w}q$-generic prescription,
{
\begin{subequations}\label{wq_generic_calibration}
\begin{align}
    \text{(}\mathrm{w}q\text{-generic)}\qquad
    \bar{\mathcal{C}}_c^{\rm generic}(\mathrm{w})
    &= 0.262285 + 0.251647\,\arctan(1.82834\,\mathrm{w}^{0.984928}),\label{Cc_generic_numeric}\\[4pt]
    \alpha_{\rm generic}(\mathrm{w})
    &= 25261.6 - 16081.8\,\arctan(363647\,\mathrm{w}^{2.09818}).\label{alpha_generic_numeric}
\end{align}
\end{subequations}
}
Both calibrations enter Eq.~\eqref{dc_gen} throughout $1/3\leq\mathrm{w}\leq1$.
The function $p(\mathrm{w},q)$ is defined by
{
\begin{equation}
    p(\mathrm{w},q) = \frac{3(1+q)}{\alpha(\mathrm{w})(2q-3)\!\left[3+\alpha(\mathrm{w})(\alpha(\mathrm{w})-3)\right]},
    \label{pfunc}
\end{equation}
}
and the $F_1(q)$, $F_2(q,\mathrm{w})$ are the Gauss hypergeometric functions
{
\begin{align}
    F_1(q) &= {}_2F_1\!\left[1,\;1-\frac{5}{2(1+q)},\;2-\frac{5}{2(1+q)},\;-q\right],\label{F1}\\[4pt]
    F_2(q,\mathrm{w}) &= {}_2F_1\!\left[1,\;1-\frac{5}{2(1+q)},\;2-\frac{5}{2(1+q)},\;-q\!\left[1-\alpha(\mathrm{w})\right]^{-2(1+q)}\right].\label{F2}
\end{align}
}
 We later denote the amplitudes obtained from Eq.~\eqref{dc_gen} with the two calibrations by $\mu_{\rm th}^{\rm fit}$ and $\mu_{\rm th}^{\rm generic}$.

%---------------------
\begin{figure}[htbp]
    \centering
    \includegraphics[width=0.98\linewidth]{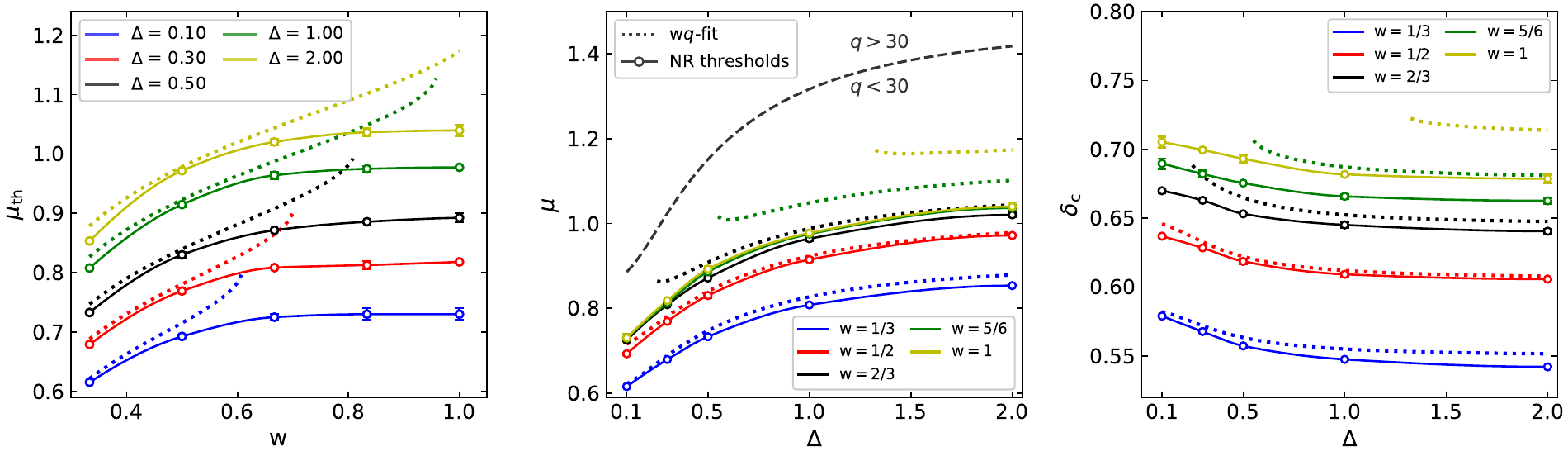}
    \caption{{NR thresholds compared with the $\mathrm{w}q$-fit prescription of Eq.~\eqref{wq_fit_calibration}. The left and middle panels show $\mu_{\rm th}$ versus $\mathrm{w}$ at fixed $\Delta$ and versus $\Delta$ at fixed $\mathrm{w}$, respectively. The right panel shows $\delta_{\rm c}=\mathcal{C}(r_m;\mu_{\rm th})$, evaluated at each method's own threshold amplitude.  Solid curves and open markers show NR results, with vertical bars indicating the amplitude intervals and their corresponding compaction ranges. Dotted curves show the $\mathrm{w}q$-fit prescription thresholds. The gray dashed curve in the middle panel marks the amplitude for which $q=30$ at each $\Delta$.}}
    \label{fig:mu_w_fit}
    %%%%%
    \vspace*{4mm}
    %%%%%
    \includegraphics[width=0.98\linewidth]{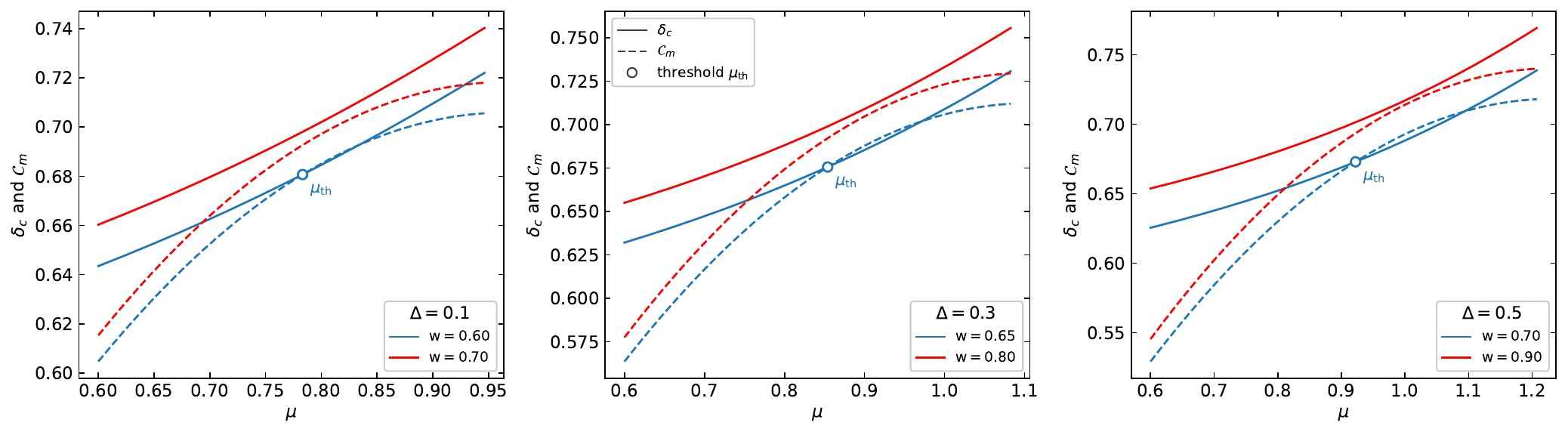}
    \caption{{Examples of crossing and non-crossing curves for the $\mathrm{w}q$-fit prescription of Eq.~\eqref{wq_fit_calibration}. The panels show $\Delta=0.1$, $0.3$, and $0.5$ from left to right. Solid curves give $\delta_c$ and dashed curves give the peak compaction $\mathcal{C}_m$ as functions of $\mu$, and the circle marks the first admissible crossing that determines the estimate for $\mu_{\rm th}$ for the $\mathrm{w}q$-fit  prescription. }}
    \label{fig:delta_compaction_previous} 
    %%%%%
    \vspace*{4mm}
    %%%%%
    \includegraphics[width=0.98\linewidth]{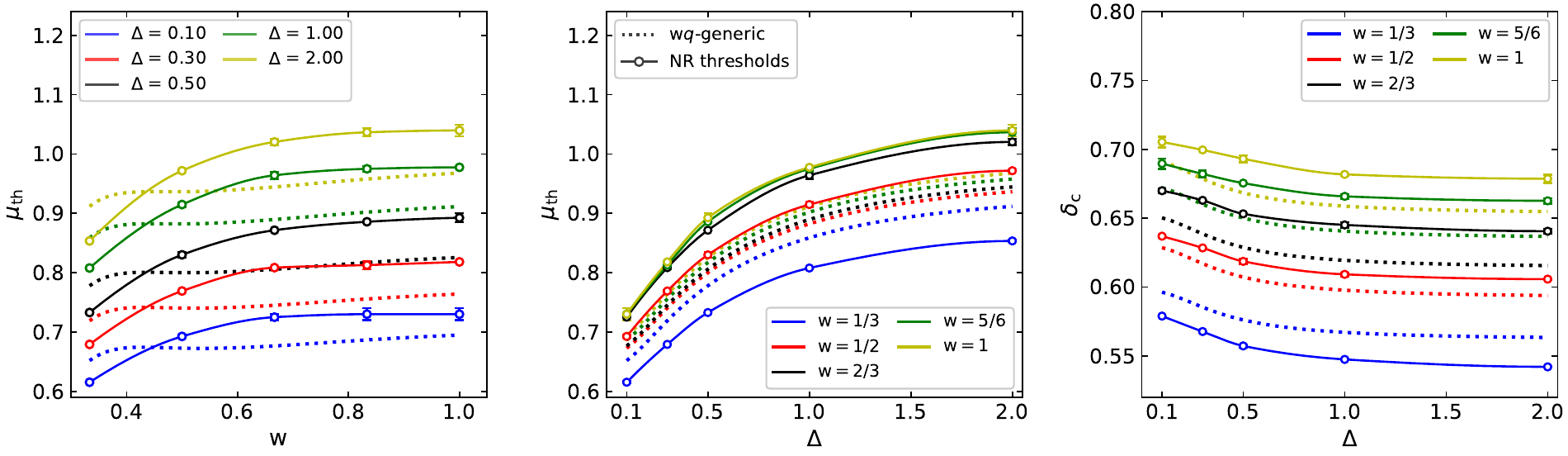}
    \caption{{NR thresholds compared with the $\mathrm{w}q$-generic prescription of Eq.~\eqref{wq_generic_calibration}, using the same profiles, colors, and NR results as in Fig.~\ref{fig:mu_w_fit}. Dotted curves show the $\mathrm{w}q$-generic prescription, while solid curves and open markers show NR. The left and middle panels display $\mu_{\rm th}$ over $\rm w$  and $\Delta$, respectively, and the right panel displays the peak compaction $\delta_c$ over $\Delta$.}}
    \label{fig:mu_w_generic}
\end{figure}
%---------------------

\subsection{{Collapse thresholds from numerical relativity versus semi-analytical prescriptions}}
\label{sec:NR_thresholds}

%---------------------
\begin{table}[t]
\begingroup

\centering
\small
\setlength{\tabcolsep}{3pt}
\caption{{Collapse amplitudes and peak compactions from NR and the $\mathrm{w}q$-fit and $\mathrm{w}q$-generic prescriptions. The column $q^{\rm NR}$ gives the shape parameter evaluated at $\mu_{\rm th}^{\rm NR}$. The columns $\epsilon_\mu^{X}$ and $\epsilon_\delta^{X}$ give percentage differences relative to NR, with $X=\mathrm{fit},\mathrm{generic}$. NR amplitudes are the midpoints of the numerical threshold intervals, whose amplitude uncertainties are below $0.01$.  Dashes indicate that the $\mathrm{w}q$-fit criterion provides no solution, as no intersection is found.}}
\label{tab:NR_thresholds}
\begin{tabular}{ccccccccccccc}
\toprule
 & & & \multicolumn{5}{c}{Profile amplitude} & \multicolumn{5}{c}{Peak compaction} \\
\cmidrule(lr){4-8}\cmidrule(lr){9-13}
$\Delta$ & $\mathrm{w}$ & $q^{\rm NR}$ & $\mu_{\rm th}^{\rm NR}$ & $\mu_{\rm th}^{\rm fit}$ & $\mu_{\rm th}^{\rm generic}$ & $\epsilon_{\mu}^{\rm fit}$ & $\epsilon_{\mu}^{\rm generic}$ & $\delta_{\rm c}^{\rm NR}$ & $\delta_{\rm c}^{\rm fit}$ & $\delta_{\rm c}^{\rm generic}$ & $\epsilon_{\delta}^{\rm fit}$ & $\epsilon_{\delta}^{\rm generic}$ \\
\midrule
0.1 & $1/3$ & 5.46 & 0.62 & 0.62 & 0.65 & $+1.00$ & $+5.93$ & 0.58 & 0.58 & 0.60 & $+0.53$ & $+2.99$ \\
0.1 & $1/2$ & 7.44 & 0.69 & 0.72 & 0.67 & $+3.30$ & $-2.85$ & 0.64 & 0.65 & 0.63 & $+1.39$ & $-1.30$ \\
0.1 & $2/3$ & 8.68 & 0.72 & \textemdash & 0.68 & \textemdash & $-6.69$ & 0.67 & \textemdash & 0.65 & \textemdash & $-2.94$ \\
0.1 & $5/6$ & 8.90 & 0.73 & \textemdash & 0.69 & \textemdash & $-5.95$ & 0.69 & \textemdash & 0.67 & \textemdash & $-2.55$ \\
0.1 & $1$ & 8.90 & 0.73 & \textemdash & 0.69 & \textemdash & $-4.83$ & 0.71 & \textemdash & 0.69 & \textemdash & $-2.04$ \\
\midrule
0.3 & $1/3$ & 4.41 & 0.68 & 0.69 & 0.72 & $+1.35$ & $+5.87$ & 0.57 & 0.57 & 0.59 & $+0.74$ & $+3.11$ \\
0.3 & $1/2$ & 5.94 & 0.77 & 0.78 & 0.74 & $+1.51$ & $-3.73$ & 0.63 & 0.63 & 0.62 & $+0.69$ & $-1.81$ \\
0.3 & $2/3$ & 6.92 & 0.81 & 0.86 & 0.75 & $+6.97$ & $-7.84$ & 0.66 & 0.68 & 0.64 & $+2.67$ & $-3.67$ \\
0.3 & $5/6$ & 7.05 & 0.81 & \textemdash & 0.76 & \textemdash & $-7.03$ & 0.68 & \textemdash & 0.66 & \textemdash & $-3.23$ \\
0.3 & $1$ & 7.21 & 0.82 & \textemdash & 0.76 & \textemdash & $-6.63$ & 0.70 & \textemdash & 0.68 & \textemdash & $-2.99$ \\
\midrule
0.5 & $1/3$ & 3.69 & 0.73 & 0.75 & 0.78 & $+1.92$ & $+6.15$ & 0.56 & 0.56 & 0.58 & $+1.09$ & $+3.39$ \\
0.5 & $1/2$ & 4.85 & 0.83 & 0.84 & 0.80 & $+1.15$ & $-3.62$ & 0.62 & 0.62 & 0.61 & $+0.56$ & $-1.85$ \\
0.5 & $2/3$ & 5.55 & 0.87 & 0.91 & 0.81 & $+4.20$ & $-7.55$ & 0.65 & 0.66 & 0.63 & $+1.80$ & $-3.73$ \\
0.5 & $5/6$ & 5.83 & 0.89 & \textemdash & 0.82 & \textemdash & $-7.77$ & 0.68 & \textemdash & 0.65 & \textemdash & $-3.75$ \\
0.5 & $1$ & 5.97 & 0.89 & \textemdash & 0.83 & \textemdash & $-7.50$ & 0.69 & \textemdash & 0.67 & \textemdash & $-3.56$ \\
\midrule
1 & $1/3$ & 3.12 & 0.81 & 0.83 & 0.86 & $+2.35$ & $+6.31$ & 0.55 & 0.55 & 0.57 & $+1.37$ & $+3.59$ \\
1 & $1/2$ & 4.02 & 0.91 & 0.92 & 0.88 & $+0.89$ & $-3.55$ & 0.61 & 0.61 & 0.60 & $+0.45$ & $-1.89$ \\
1 & $2/3$ & 4.60 & 0.96 & 0.99 & 0.89 & $+2.45$ & $-7.73$ & 0.65 & 0.65 & 0.62 & $+1.13$ & $-3.98$ \\
1 & $5/6$ & 4.75 & 0.97 & 1.05 & 0.90 & $+7.55$ & $-7.50$ & 0.67 & 0.69 & 0.64 & $+3.21$ & $-3.79$ \\
1 & $1$ & 4.78 & 0.98 & \textemdash & 0.91 & \textemdash & $-6.77$ & 0.68 & \textemdash & 0.66 & \textemdash & $-3.38$ \\
\midrule
2 & $1/3$ & 2.89 & 0.85 & 0.88 & 0.91 & $+2.96$ & $+6.83$ & 0.54 & 0.55 & 0.56 & $+1.75$ & $+3.94$ \\
2 & $1/2$ & 3.75 & 0.97 & 0.98 & 0.94 & $+0.68$ & $-3.64$ & 0.61 & 0.61 & 0.59 & $+0.35$ & $-1.96$ \\
2 & $2/3$ & 4.22 & 1.02 & 1.04 & 0.94 & $+2.33$ & $-7.43$ & 0.64 & 0.65 & 0.62 & $+1.10$ & $-3.90$ \\
2 & $5/6$ & 4.41 & 1.04 & 1.10 & 0.96 & $+6.29$ & $-7.61$ & 0.66 & 0.68 & 0.64 & $+2.77$ & $-3.91$ \\
2 & $1$ & 4.44 & 1.04 & 1.17 & 0.97 & $+12.87$ & $-6.92$ & 0.68 & 0.71 & 0.65 & $+5.19$ & $-3.52$ \\
\bottomrule
\end{tabular}
\endgroup
\end{table}
%%%

We use NR simulations to evolve the Hubble re-entry of the curvature profiles described in Sec.~\ref{sec:peak_theory} in a universe filled with a perfect fluid, $P=\rm w\rho$. The profiles are initialized well outside the Hubble radius at $k_p/(a_{\rm ini}H_{\rm ini})=0.1$. For each pair $(\Delta,\rm w)$, the amplitude $\mu$ is varied to identify the threshold separating profiles that disperse from those that collapse to form a PBH. The outcome is set by the competition between self-gravity in full general relativity and fluid pressure rather than by a semi-analytical threshold condition. Further details of the numerical implementation and constraint monitoring are provided in the Appendix.

The NR thresholds are calculated to an accuracy of $0.01$ and shown as solid curves in Figs.~\ref{fig:mu_w_fit} and \ref{fig:mu_w_generic}. In both figures, the left panel shows $\mu_{\rm th}$ versus $\mathrm{w}$ at fixed $\Delta$, and the middle panel shows $\mu_{\rm th}$ versus $\Delta$ at fixed $\mathrm{w}$. The numerical threshold increases with both the stiffness of the background and the width of the power spectrum. At fixed $\Delta$, increasing $\mathrm{w}$ raises the pressure opposing collapse and therefore requires a larger initial curvature amplitude. The dependence becomes weaker toward $\mathrm{w}\simeq1$, particularly for the narrowest profiles. At fixed $\mathrm{w}$, broader spectra also require larger $\mu$. For the broadest case studied, $\Delta=2$, the threshold exceeds unity for $\mathrm{w}\geq2/3$.

Fig.~\ref{fig:mu_w_fit} compares the NR thresholds with the $\mathrm{w}q$-fit prescription of Eq.~\eqref{wq_fit_calibration}, shown by dotted curves. The $\mathrm{w}q$-fit prescription closely follows the NR results only for narrow spectra and near radiation domination. For stiffer backgrounds, the prescription cannot always provide a threshold because the peak compaction $\mathcal{C}_m$ does not reach $\delta_c$, so the condition $\mathcal{C}_m\geq\delta_c$ cannot be satisfied. Fig.~\ref{fig:delta_compaction_previous} illustrates this by plotting $\mathcal{C}_m$ and $\delta_c$ against $\mu$ for different spectral widths. The threshold exists only when the two curves cross. At smaller $\mathrm{w}$, they intersect at the open circle. As $\mathrm{w}$ increases, the $\mathrm{w}q$-fit value of $\delta_c$ remains above $\mathcal{C}_m$ and the curves no longer meet. The range of $\mathrm{w}$ for which the prescription provides a threshold broadens as $\Delta$ increases, although the discrepancy with NR also grows.

% A comparison between the NR results and the $\mathrm{w}q$-generic prescription of Eq.~\eqref{wq_generic_calibration} is provided in Fig.~\ref{fig:mu_w_generic}. 
For all cases studied, the shape parameter evaluated at the NR threshold lies in the range $2<q^{\rm NR}<9$, within the $q$ range used to calibrate the $\mathrm{w}q$-fit prescription. For completeness, we also compare the NR results with the $\mathrm{w}q$-generic prescription of Eq.~\eqref{wq_generic_calibration}, as shown in Fig.~\ref{fig:mu_w_generic}.
This prescription is built to satisfy the condition $\delta_c\to f(\mathrm{w})$ as $q\to\infty$, and a threshold can be found in all cases considered here, including stiff backgrounds where the $\mathrm{w}q$-fit prescription fails. The dotted curves show the thresholds obtained from the $\mathrm{w}q$-generic prescription, while the solid lines show the NR results. The resulting thresholds lie above the NR values for the sampled $\mathrm{w}<1/2$ and below them for the sampled $\mathrm{w}\geq1/2$. Moreover, the difference from the NR results tends to grow as the spectrum broadens.

The right panels of Figs.~\ref{fig:mu_w_fit} and \ref{fig:mu_w_generic} compare the peak compaction $\delta_c$ with the NR value $\delta_c^{\rm NR}$ as a function of $\Delta$, with each quantity evaluated at the corresponding method's threshold. Solid curves show NR and dotted curves show the prescriptions. Both prescriptions give smaller relative discrepancies in peak compaction than in the amplitude threshold.

The results are collected in Table~\ref{tab:NR_thresholds} which shows the resulting threshold for NR and both prescriptions, and also the corresponding peak. The columns $\epsilon_\mu$ and $\epsilon_\delta$ show the relative differences of the prescriptions with the NR ones. For the $wq$-fit prescription, the agreement is best near the radiation domination ${\rm w}$ where the relative differences are about $\epsilon_\mu^{\rm fit} = 1\text{--}3\%$. For cases in which a threshold exists, the relative differences grow with increasing $\rm w$ and $\Delta$ so that they reach about $12.9\%$ for $\Delta = 2$ and ${\rm w} = 1$. The $wq$-generic prescription gives a solution in all cases. It overestimates the threshold for the radiation case with the relative differences about $\epsilon_\mu^{\rm generic} = 5.9\text{--}6.8\%$, and underestimates the threshold for the stiffer background.  Its best agreement occurs at $\mathrm{w}=1/2$, with a relative difference of $|\epsilon_\mu^{\rm generic}|=2.8\text{--}3.7\%$, and the discrepancy remains within about $7.8\%$ overall. In general, the relative differences in peak compaction are smaller for both prescriptions.  For the $\mathrm{w}q$-fit prescription, $\epsilon_\delta$ is very small near radiation domination. It grows with $\mathrm{w}$ and $\Delta$, with $\epsilon_\delta^{\rm fit}=5\%$ at $\mathrm{w}=1$ and $\Delta=2$. For the $wq$-generic prescription, the differences near radiation domination are larger than those of the $wq$-fit prescription, but remain within $4\%$ overall.

The larger relative discrepancies in $\mu_{\rm th}$ compared with those in $\delta_c$ are partly due to the nonlinear relation between the profile amplitude and the peak compaction, and therefore moderate relative differences in $\delta_c$ translate into larger relative differences in $\mu_{\rm th}$.

Since the PBH abundance is exponentially sensitive to the threshold, we use the NR thresholds in Table~\ref{tab:NR_thresholds} for all subsequent calculations of the abundance and induced gravitational-wave spectrum.

\section{PBH abundance}
\label{sec:abundance}

\begin{figure}[b]
    \centering
    \begin{minipage}[c]{0.4\linewidth}
        \centering
        \includegraphics[width=0.9\linewidth]{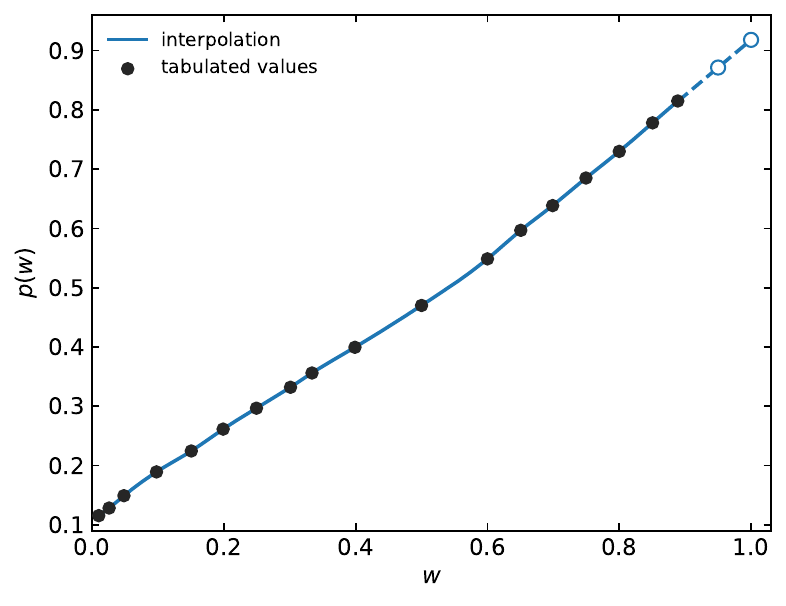}
    \end{minipage}\hspace{0.025\linewidth}%
    \begin{minipage}[c]{0.4\linewidth}
        \centering
        \small
        \renewcommand{\arraystretch}{1.08}
        \setlength{\tabcolsep}{4pt}
        \begin{tabular}{cc@{\hspace{0.65em}}|@{\hspace{0.5em}}cc}
            \toprule
            $\rm w$ & $p(\rm w)$ & $\rm w$ & $p(\rm w)$ \\
            \midrule
            0.010 & 0.116 & 0.500 & 0.470 \\
            0.026 & 0.128 & 0.600 & 0.549 \\
            0.048 & 0.149 & 0.651 & 0.597 \\
            0.098 & 0.189 & 0.699 & 0.639 \\
            0.151 & 0.225 & 0.749 & 0.685 \\
            0.199 & 0.261 & 0.800 & 0.730 \\
            0.249 & 0.297 & 0.851 & 0.778 \\
            0.301 & 0.332 & 0.889 & 0.815 \\
            0.334 & 0.356 & $0.950^{\ast}$ & 0.871 \\
            0.399 & 0.399 & $1.000^{\ast}$ & 0.918 \\
            \bottomrule
        \end{tabular}
    \end{minipage}
    \caption{Dependence of the critical exponent $p(\rm w)$ on the equation of state. The points show the values reported in Refs.~\citep{Maison:1995cc}, and the solid curve shows their interpolation. The dashed segment and the open marker show the extrapolation toward $w=1$. The table on the right lists the same values together with the two extrapolated entries, which are marked by asterisks.}
    \label{fig:p_w}
\end{figure}

When the peak amplitude $\mu$ exceeds the threshold $\mu_{\rm th}$, a PBH forms with mass obeying a universal power-law relation given as \citep{Choptuik:1992jv,Niemeyer:1997mt,Inui:2024fgk},
\begin{equation}\label{mass_scaling}
    M_{\rm PBH} = M_H \; \mathcal{K}\,(\mu-\mu_{\rm th})^{p(\rm w)},
\end{equation}
{The dimensionless coefficient $\mathcal{K}$ sets the mass normalization. We use the representative value $\mathcal{K}\simeq4.36$ found for profiles with $q>4$ in Ref.~\citep{Escriva:2021pmf}.}
The critical exponent depends on the background equation-of-state parameter. We denote this dependence by $p(\rm w)$ and interpolate the values obtained from analytical and numerical studies~\citep{Maison:1995cc} (see also \citep{Musco:2012au,Joana:2025gqf}). A linear extrapolation of the highest-$\rm w$ values gives $p(1)=0.918$. The interpolation and the values used in our calculations are shown in Fig.~\ref{fig:p_w}. 
% The exponent parameter $p$ is a universal exponent that, in general, depends on the equation of state $\rm w$. The analytical solution and numerical solution have revealed this dependency behavior~\citep{Maison:1995cc,Musco:2012au}.

%---------------------
%--------------------------

\begin{figure}
    \centering
    \includegraphics[width=0.98\linewidth]{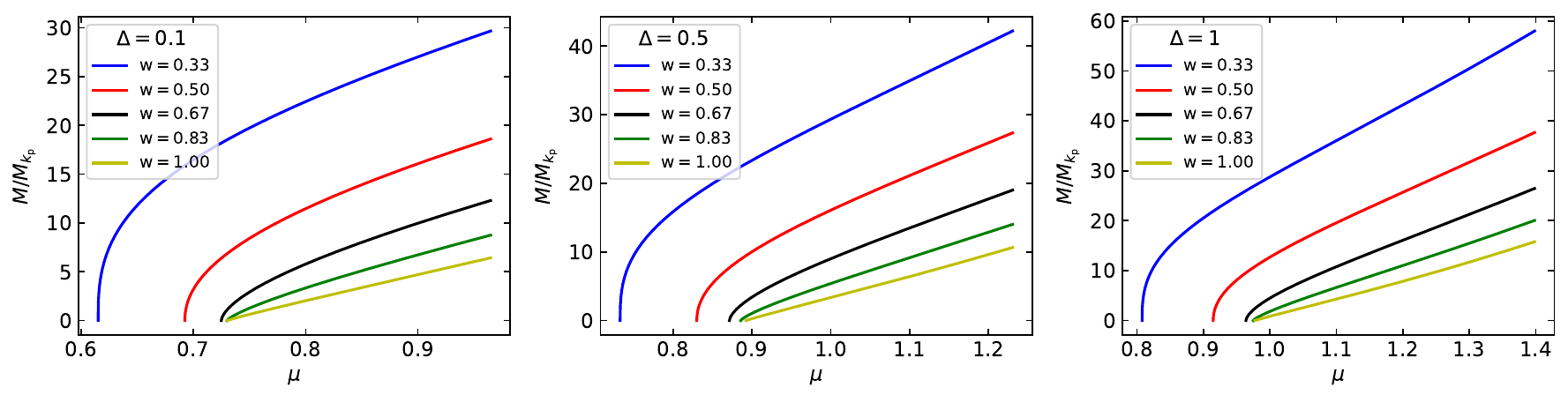}
    % \caption{The figure displays the normalized PBH mass as a function of $\mu$ for three sets of parameters. }
    \caption{{PBH mass in units of $M_{k_p}$ versus profile amplitude $\mu$. The panels show $\Delta=0.1$, $0.5$, and $1$, with curves for $\mathrm{w}=1/3$, $1/2$, $2/3$, $5/6$, and $1$.}}
    \label{fig:mass_mu}
\end{figure}

$M_H(\mu)$ stands for the mass inside the horizon at the time of re-entry, which, in the case of spherical symmetry, is given as~\citep{Escriva:2024aeo,Escriva:2024lmm}
\begin{equation}
    M_H(\mu) = \big( r_m \, e^{\zeta(r_m)} \big) ^{l(\rm w)} \, M_{k_p}
\end{equation}
Here $l(\mathrm{w})=3(1+\mathrm{w})/(1+3\mathrm{w})$, and $r_m$ is the position of the compaction maximum in units of $k_p^{-1}$. The reference mass $M_{k_p}$ is the background Hubble-horizon mass when the peak mode satisfies $k_p=aH$. We assume a constant equation of state until an instantaneous transition to radiation domination at $k_{\rm rh}=a_{\rm rh}H_{\rm rh}$. Throughout the abundance and SIGW calculations, we fix $k_p=1.56\times10^{13}\,{\rm Mpc}^{-1}$ and $k_{\rm rh}/k_p=0.01$, so that the modes around the curvature-spectrum peak re-enter before reheating. Matching the horizon mass to the subsequent radiation era gives~\citep{Domenech:2021ztg}
\begin{equation}\label{Mkstar}
    M_{k_p}=10^{20}\!\left(\frac{g_\star}{106.75}\right)^{-1/6}
    \!\left(\frac{k_p}{1.56\times10^{13}\,{\rm Mpc}^{-1}}\right)^{-2}
    \!\left(\frac{k_{\rm rh}}{k_p}\right)^b\,{\rm g},
    \qquad b=\frac{1-3\mathrm{w}}{1+3\mathrm{w}}.
\end{equation}
Here $g_\star$ is the effective number of relativistic energy degrees of freedom at reheating. We take the energy and entropy degrees of freedom to be equal, $g_\star=g_{s,\rm rh}=106.75$. With this choice of reheating scale, $M_{k_p}$ ranges from $10^{20}\,{\rm g}$ at $\mathrm{w}=1/3$ to $10^{21}\,{\rm g}$ at $\mathrm{w}=1$. Fig.~\ref{fig:mass_mu} shows the dimensionless PBH mass as a function of $\mu$, with the onset of each curve set by the corresponding NR threshold.

%
%---------------------
%--------------------------

Based on the peak theory, the comoving number density of PBHs at mass $M$ in a comoving volume is obtained as ~\citep{Pi:2024ert,Inui:2024fgk}
\begin{equation}\label{N_pbh}
    \mathcal{N}_{\rm PBH} (M) = \int dK\int d\mu\;\delta_D\!\left(\ln\frac{M}{M(\mu,K)}\right)\mathcal{N}_{\rm pk}(\mu,K).
\end{equation}
Here $\delta_D$ is the Dirac delta distribution, $M(\mu,K)$ is the PBH mass associated with a peak of height $\mu$ and curvature $K$, and $\mathcal{N}_{\rm pk}(\mu,K)$ is the comoving peak number density per unit $\mu$ and $K$, given by
\begin{equation}\label{Npk}
    \mathcal{N}_{\rm pk}(\mu,K)\,d\mu\,dK = 2\!\left(\frac{1}{6\pi}\right)^{\!3/2}\!\frac{\sigma_2^2}{\sigma_1^4}\frac{\sigma_4^3}{\sigma_3^3}\,\mu K\, f\!\left(\frac{\sigma_2}{\sigma_1^2}\mu K^2\right)P_1^{(3)}\!\left(\frac{\sigma_2}{\sigma_1^2}\mu,\frac{\sigma_2}{\sigma_1^2}\mu K^2\right)d\mu\,dK,
\end{equation}
with the auxiliary functions
\begin{align}
    f(x) &= \frac{x^3-3x}{2}\!\left[{\rm erf}\!\left(\sqrt{\tfrac{5}{2}}\,x\right)+{\rm erf}\!\left(\tfrac{1}{2}\sqrt{\tfrac{5}{2}}\,x\right)\right]\nonumber\\
    &\quad+\sqrt{\frac{2}{5\pi}}\!\left[\!\left(\frac{31x^2}{4}+\frac{8}{5}\right)\!e^{-5x^2/8}+\!\left(\frac{x^2}{2}-\frac{8}{5}\right)\!e^{-5x^2/2}\right],\\[4pt]
    P_1^{(3)}(v,x) &= \frac{1}{2\pi\sqrt{1-\gamma_3^2}}\exp\!\left[-\frac{1}{2}\!\left(v^2+\frac{(x-\gamma_3 v)^2}{1-\gamma_3^2}\right)\right].
\end{align}
We normalize the present scale factor to $a_0=1$, so the comoving number density relates to the present physical number density. TWe normalize the present scale factor to $a_0=1$, so the comoving number density equals the present physical number density. The PBH mass function gives the fraction of present dark matter per logarithmic mass interval~\citep{Escriva:2024lmm,Escriva:2024aeo},
\begin{equation}\label{fpbh}
    f_{\rm PBH}(M) = \frac{M\,\mathcal{N}_{\rm PBH}(M)}{\rho_{\rm DM,0}},
    \qquad \rho_{\rm DM,0}=3M_p^2H_0^2\Omega_{\rm DM,0}.
\end{equation}
Here $\rho_{\rm DM,0}$ is the present dark-matter density, $M_p=(8\pi G)^{-1/2}$ is the reduced Planck mass, $H_0$ is the present Hubble parameter, and $\Omega_{\rm DM,0}$ is the dark-matter density in units of the present critical density. 

The total PBH abundance is obtained by integrating the PBH mass function as
\begin{equation}\label{fpbh_tot}
    f_{\rm PBH}^{\rm tot} = \int f_{\rm PBH}(M)\,d\ln M.
\end{equation}
%---------------------------------------
\begin{figure}[t]
    \centering
    \includegraphics[width=0.98\linewidth]{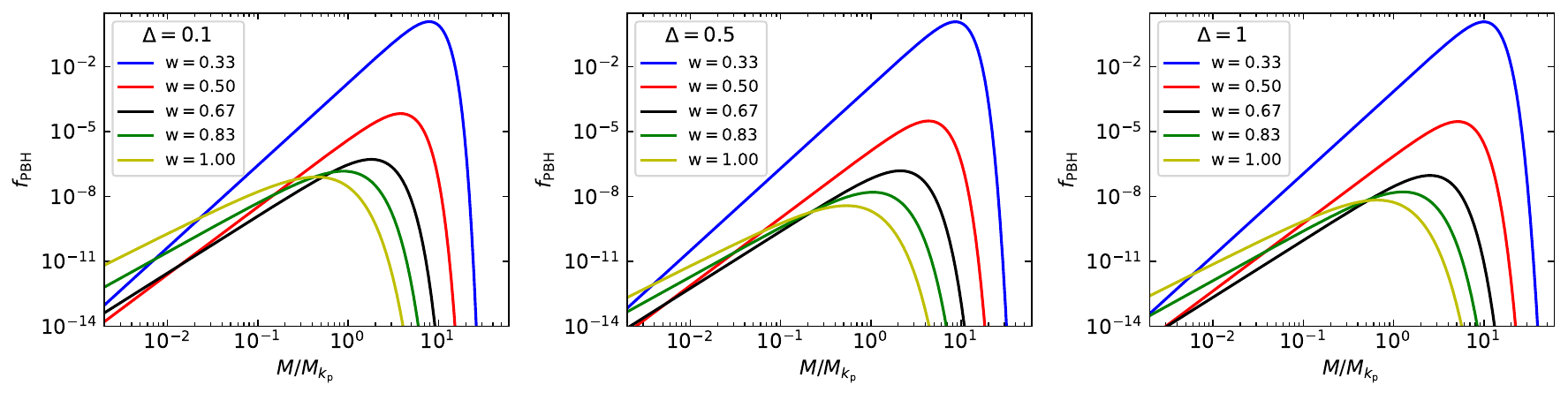}
    % \caption{The figure shows the behavior of $f_{\rm PBH}$ versus the mass $M_{PBH}/M_{k_{\rm p}}$ for different values of $\rm w$ and $\Delta$, and exhibits the dependency of the PBH mass function on the equation of state parameter $\rm w$ and the width parameter $\Delta$.}
    \caption{{PBH mass functions for $\Delta=0.1$, $0.5$, and $1$, with curves for $\mathrm{w}=1/3$, $1/2$, $2/3$, $5/6$, and $1$. In each panel, $A$ is fixed to give $f_{\rm PBH}^{\rm tot}=1$ at $\mathrm{w}=1/3$ and held constant as $\mathrm{w}$ varies, using the NR thresholds and $k_{\rm rh}/k_p=0.01$. Dashed tails extrapolate the mass-scaling relation beyond the type-I boundary.}}
    \label{fig:fpbh}
\end{figure}
%------------------------
{Fig.~\ref{fig:fpbh} compares the PBH mass functions at fixed primordial amplitude within each panel. For each width, we choose $A$ such that $f_{\rm PBH}^{\rm tot}=1$ in radiation domination and retain that value for the other equations of state. The corresponding amplitudes are $A\simeq0.01441$, $0.03064$, and $0.05468$ for $\Delta=0.1$, $0.5$, and $1$, respectively. Increasing $\mathrm{w}$ raises the collapse threshold, so fewer peaks of the same primordial spectrum form PBHs. The total abundance is consequently suppressed by several orders of magnitude, while the mass-function peak shifts toward smaller $M/M_{k_p}$. Only the radiation curves have unit total abundance. The dashed tails continue the mass-scaling relation and peak-density expression beyond the type-I boundary for illustration only. They are not predictions for type-II collapse and are excluded from the total abundance integral. Fig.~\ref{fig:ftot} shows how the total abundance depends on $A$. For the adopted reheating scale, a larger amplitude is needed to recover $f_{\rm PBH}^{\rm tot}=1$ as either $\Delta$ or $\mathrm{w}$ increases.}
%----------------
\begin{figure}
    \centering
    \includegraphics[width=0.98\linewidth]{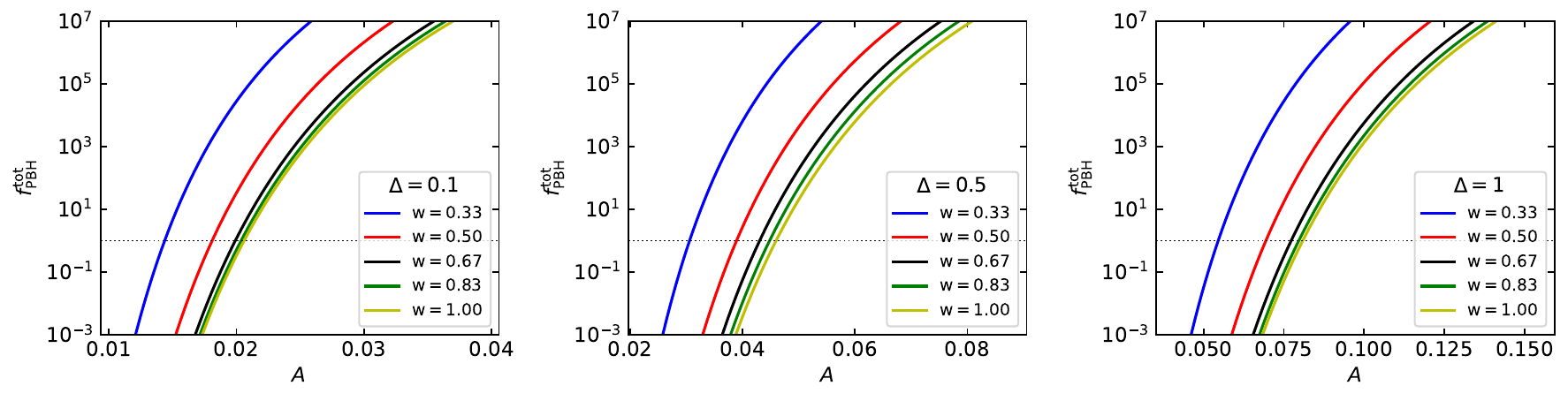}
    % \caption{The figure displays the total PBH abundance $f_{\rm PBH}^{\rm tot}$ versus $A$ for different values of equation of state parameter $ \rm w$.  }
    \caption{{Total PBH abundance versus integrated curvature-spectrum amplitude $A$ for the same parameters as Fig.~\ref{fig:fpbh}. The horizontal dotted line marks $f_{\rm PBH}^{\rm tot}=1$.}}
    \label{fig:ftot}
\end{figure}
%----------------

\begin{figure}[!h]
    \centering
    \includegraphics[width=0.45\linewidth]{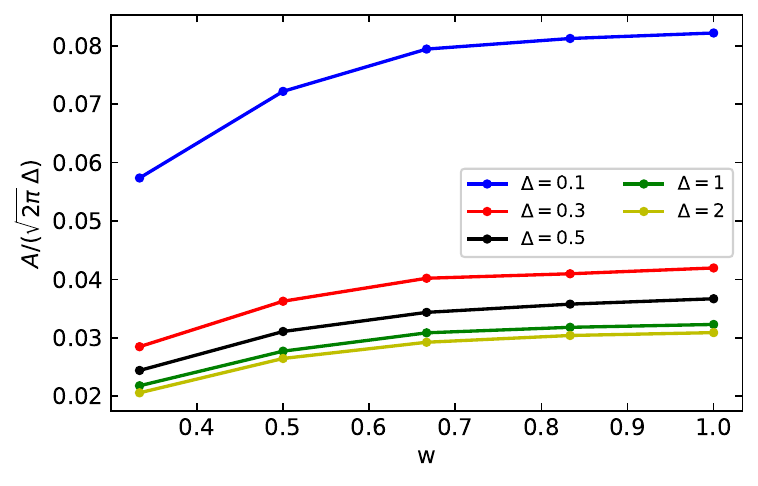}
    \caption{{Curvature-spectrum peak amplitude $A/(\sqrt{2\pi}\Delta)$ required for $f_{\rm PBH}^{\rm tot}=1$, using the NR thresholds, $k_p=1.56\times10^{13}\,{\rm Mpc}^{-1}$, and $k_{\rm rh}/k_p=0.01$.}}
    \label{fig:Aw_ftot}
\end{figure}

{Fig.~\ref{fig:Aw_ftot} shows the peak amplitude $A/(\sqrt{2\pi}\Delta)$ required for PBHs to account for all of the dark matter. At fixed $\Delta$, the peak amplitude increases with $\mathrm{w}$. At fixed $\mathrm{w}$, however, broader spectra require a lower peak amplitude even though their integrated amplitude $A$ is larger. The distinction matters for the related SIGW signal, which depends on the power over a range of wavenumbers and not only on its peak value.}
%---------------
%------------------
%Taken together, both a wider spectrum and a stiffer background push the required amplitude upward, so reaching the full dark matter fraction becomes harder in either direction.

%------------------------------------------
%------------------------------------------
%--------------     6. SIGW     
%------------------------------------------
%------------------------------------------
\section{Scalar-Induced Gravitational Waves}
\label{sec:sigw}

The cosmological perturbations generated during inflation are divided into scalar, vector, and tensor perturbations. The scalar and tensor perturbations are usually of interest to us, as the vector perturbations are diluted due to the extreme expansion. The scalar perturbations are the seeds to the universe's structure, and the tensor perturbations are the source of the primordial gravitational waves. At the first order in perturbation theory, the scalar and tensor perturbations are evolved independently, and the gravitational waves propagate source-free (if we assume that there are no anisotropic stresses). Then, the tensor power spectrum is related to the primordial power spectrum through the transfer function $T(k,\eta)$ \citep{Caprini:2018mtu,Christensen:2018iqi},
\begin{equation}
    \Delta_h(k,\eta) = T(k,\eta)\,\mathcal{P}_{h,{\rm inf}}(k),
    \label{Delta_h}
\end{equation}
where $\eta$ is conformal time, and 
\begin{equation}
    \mathcal{P}_{h,{\rm inf}}(k) = \frac{2}{\pi^2}\,\frac{H_\star^2}{M_p^2}\left(\frac{k}{k_\star}\right)^{n_t}
    \label{Ph_inf}
\end{equation}
is the primordial tensor power spectrum, with $H_\star$ the Hubble parameter at the crossing time, and $n_t$ the tensor spectral index. The transfer function $T(k,\eta)$ contains information about the background evolution from the time the mode $k$ re-enters the horizon to the time of observations. Assuming that the thermal history of the universe at the post-inflationary time and before the radiation-dominant phase includes a non-standard epoch with an equation of state $\rm w$ different from that of radiation, the transfer function changes from its standard form. In this case, the first-order gravitational waves energy density parameter today becomes~\citep{Figueroa:2019paj,Bernal:2019lpc,Bernal:2020ywq}
\begin{equation}\label{OmegaGW1}
    \Omega_{{\rm GW},0}^{(1)}(k) = \frac{\Omega_{{\rm rad},0}}{12\pi^2}\left(\frac{g_{*,k}}{g_{s,k}}\right)\!\left(\frac{g_{s,0}}{g_{s,k}}\right)^{\!4/3}\!\left(\frac{H_\star}{M_p}\right)^{\!2}\frac{\Gamma^2\!\left(\alpha+\tfrac{1}{2}\right)}{2^{2(1-\alpha)}\alpha^{2\alpha}\Gamma^2\!\left(\tfrac{3}{2}\right)}\,\mathcal{W}(\kappa)\,\kappa^{2(1-\alpha)},
\end{equation}
where $\alpha = 2/(1+3\rm w)$, $\kappa = k/k(T_1)$, $T_1$ is the temperature at the start of the radiation phase, and $g_{*,k}$, $g_{s,k}$ are the relativistic and entropy degrees of freedom at the re-entry of mode $k$. In addition, the spectral window function is given by
\begin{equation}
    \mathcal{W}(\kappa) = \frac{\pi\alpha}{2\kappa}\!\left[\left(\kappa\,J_{\alpha+1/2}(\kappa) - J_{\alpha-1/2}(\kappa)\right)^2 + \kappa^2\,J^2_{\alpha-1/2}(\kappa)\right],
    \label{Wkappa}
\end{equation}
where $J_i$ is the Bessel function of order $i$. At $\rm w = 1/3$, Eqs.~\eqref{OmegaGW1} and \eqref{Wkappa} reduce to the standard radiation-dominated expressions. \\

At the second order of the perturbation theory, the tensor and scalar perturbations are coupled, and they do not evolve independently. The scalar perturbations act as the source term in the evolution equation of the tensor perturbations and lead to scalar-induced gravitational waves (SIGW) \citep{Ananda:2006af,Baumann:2007zm,Kohri:2018awv,Espinosa:2018eve,Domenech:2021ztg}. The large enhancement in the scalar power spectrum generates a gravitational wave signal that may fall within the sensitivity bounds of present and future detectors. %Products of first-order scalar Bardeen potentials act as a classical source for second-order tensor fluctuations. 
The evolution equation for the tensor polarization $h_k$ in conformal time is~\citep{Ananda:2006af,Baumann:2007zm,Kohri:2018awv,Espinosa:2018eve}
\begin{equation}\label{hk_eq}
    h_k'' + 2\mathcal{H}\,h_k' + k^2 h_k = \mathcal{S}(\mathbf{k},\eta),
\end{equation}
Here primes denote derivatives with respect to $\eta$,  $\mathcal{H}=aH$ is the conformal Hubble parameter and $\mathcal{S}(\mathbf{k},\eta)$ is the scalar source. The SIGW energy density per logarithmic wavenumber interval, normalized to the total energy density, is~\citep{Kohri:2018awv,Espinosa:2018eve,Domenech:2021ztg}
\begin{equation}\label{OmegaGW_def}
    \Omega_{\rm GW}(k,\eta) = \frac{k^2}{12H^2 a^2}\,\mathcal{P}_h(k,\eta).
\end{equation}
where $\mathcal{P}_h(k,\eta)$ denotes the dimensionless power spectrum of the induced tensor perturbations.
For modes well inside the horizon before reheating, the constant-$\mathrm{w}$ solution gives~\citep{Domenech:2021ztg,Domenech:2020kqm,Domenech:2024rks,Liu:2023hpw,Chen:2024roo}

\begin{equation}\label{OmegaGW_w}
    \Omega_{\rm GW}^{\rm w}(k) = \mathcal{N}\left(\frac{k}{k_{\rm rh}}\right)^{\!-2b}\!\int_0^1\!\mathrm{d}d\int_1^\infty\!\mathrm{d}s\; I_{\rm w}(d,s)\,\mathcal{P}_\zeta\!\!\left(\frac{k(s+d)}{2}\right)\!\mathcal{P}_\zeta\!\!\left(\frac{k(s-d)}{2}\right),
\end{equation}
where $b=(1-3\mathrm{w})/(1+3\mathrm{w})$. The integration variables are $s=(k_1+k_2)/k$ and $d=|k_1-k_2|/k$, where $k_1$ and $k_2$ are the magnitudes of the two scalar wavevectors sourcing the tensor mode $k$. Symmetry under their exchange allows the integration to be restricted to $0\leq d\leq1$. The kernel $I_{\mathrm{w}}(d,s)$ is expressed in terms of Legendre and Ferrers functions, with sound speed $c_s^2=\mathrm{w}$ for the adiabatic perfect fluid considered here~\citep{Domenech:2021ztg,Liu:2023hpw}. We set $\mathcal{N}=1$ to evaluate the spectrum at the transition to radiation domination and then apply Eq.~\eqref{OmegaGW_today}. The factor $(k/k_{\rm rh})^{-2b}$ accounts for the different redshifting of the GWs and the background fluid before reheating. It enhances the GW energy fraction for $\mathrm{w}>1/3$ and becomes unity in radiation domination.

After reheating, the GW energy density redshifts as radiation. Including the change in relativistic degrees of freedom, the present spectrum is~\citep{Pi:2020otn,Kohri:2018awv}
\begin{equation}\label{OmegaGW_today}
    % \Omega_{{\rm GW},0}(f)\,h^2 = 1.6\times10^{-5}\!\left(\frac{g_{*s}(\eta_k)}{106.75}\right)^{-1/3}\!\!\left(\frac{\Omega_{{\rm r},0}h^2}{4.1\times10^{-5}}\right)\Omega_{{\rm GW,r}}(f),
    \Omega_{{\rm GW},0}(f)\,h^2 = 1.6\times10^{-5}\!\left(\frac{g_{s,\rm rh}}{106.75}\right)^{-1/3}\!\!\left(\frac{\Omega_{{\rm rad},0}h^2}{4.1\times10^{-5}}\right)\Omega_{{\rm GW,r}}(f)~,
\end{equation}
where $h\equiv H_0/(100\,{\rm km\,s^{-1}\,Mpc^{-1}})$ is the dimensionless present Hubble parameter $\Omega_{{\rm GW,r}}(f)$ is the GW energy fraction at the onset of radiation domination, obtained from Eq.~\eqref{OmegaGW_w}, and $g_{s,\rm rh}$ denotes the entropy degrees of freedom at that transition. We have written the redshift factor for equal energy and entropy degrees of freedom, as assumed in our calculation. The frequency $f$  is related to the comoving mode $k$ through the following relation 
\begin{equation}
    f = 1.546\times10^{-15}\left(\frac{k}{{\rm Mpc}^{-1}}\right){\rm Hz}.
    \label{k_to_f}
\end{equation}

We evaluate Eq.~\eqref{OmegaGW_w} with \texttt{SIGWfast}~\citep{Witkowski:2022mtg}, using the NR thresholds to fix $A$ through $f_{\rm PBH}^{\rm tot}=1$ for every $(\Delta,\mathrm{w})$ pair. The same reheating scale is used in the PBH mass normalization and in the GW calculation. Our choice $k_{\rm rh}/k_p=0.01$ gives $k_{\rm rh}=1.56\times10^{11}\,{\rm Mpc}^{-1}$ and $f_{\rm rh}\simeq2.4\times10^{-4}\,{\rm Hz}$, while the curvature-spectrum peak corresponds to $f_p\simeq2.4\times10^{-2}\,{\rm Hz}$. Reheating is instantaneous in this model, but occurs after the peak modes have re-entered.

\begin{figure}
    \centering
    \subfigure{\includegraphics[width=0.95\linewidth]{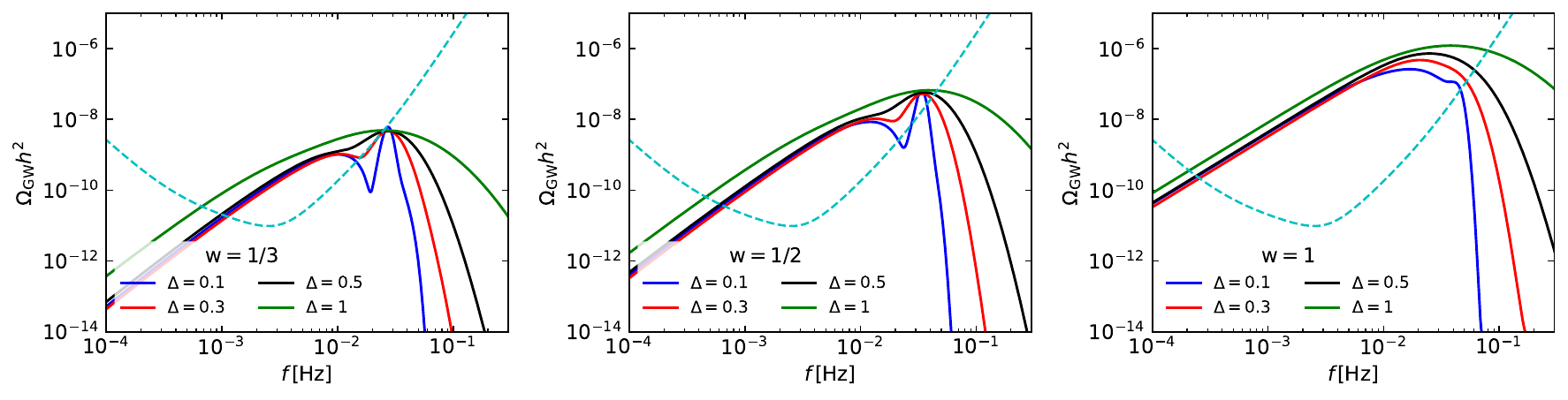}}
    \subfigure{\includegraphics[width=0.95\linewidth]{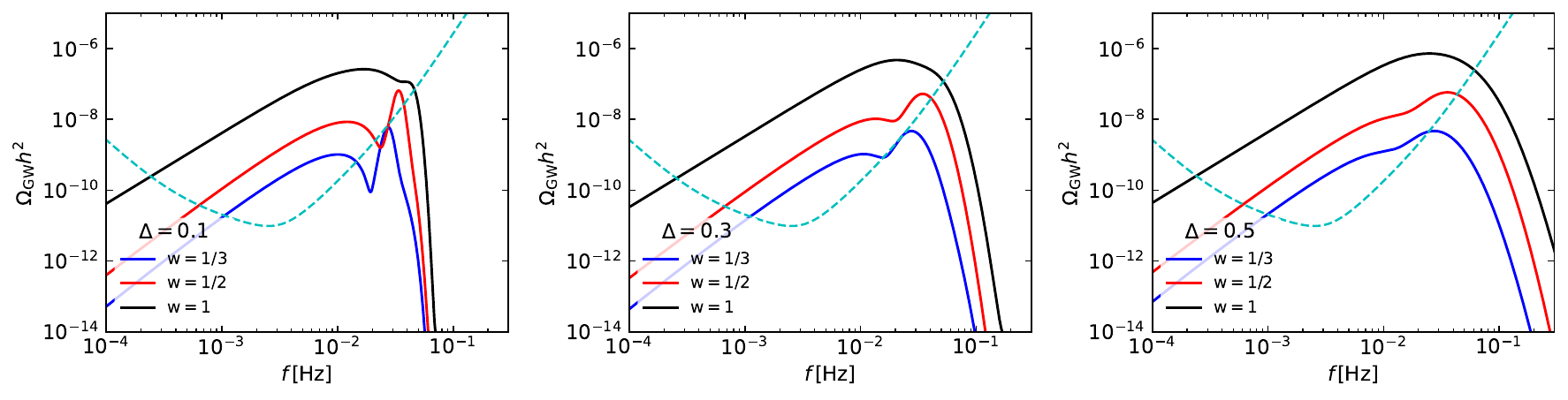}}
    % \caption{The plots show the energy spectrum of SIGW versus frequency for different sets of parameters. The dashed line represents the LISA sensitivity curve.}
    \caption{{SIGW spectra for $k_p=1.56\times10^{13}\,{\rm Mpc}^{-1}$, $k_{\rm rh}/k_p=0.01$, and $f_{\rm PBH}^{\rm tot}=1$. The upper row varies $\Delta$ at fixed $\mathrm{w}=1/3$, $1/2$, and $1$; the lower row varies $\mathrm{w}$ at fixed $\Delta=0.1$, $0.3$, and $0.5$. The cyan dashed curve shows the LISA sensitivity curve.}}
    \label{fig:sigw}
\end{figure}

Fig.~\ref{fig:sigw} shows the resulting spectra. Increasing $\Delta$ broadens the SIGW peak and extends its high-frequency tail. A smaller curvature-spectrum peak can therefore produce a stronger signal over part of the GW spectrum. Increasing $\mathrm{w}$ enhances the signal relative to radiation domination and changes the low-frequency slope. The double-peaked structure visible for narrow spectra becomes less pronounced toward the stiff limit, where a broad maximum dominates. The peak frequency depends on both parameters and on which spectral maximum is largest. The spectra are compared with the LISA sensitivity curve~\citep{Barausse:2020rsu,LISACosmologyWorkingGroup:2022jok,Bartolo:2018evs}.

The two tails provide complementary information. When the finite width of the scalar spectrum is resolved, the leading infrared dependence above reheating is $\Omega_{\rm GW}\propto k^{3-2|b|}$ for $k_{\rm rh}\ll k\ll k_p$~\citep{Domenech:2020kqm,Domenech:2021ztg}. In the stiff regime $b<0$, this is $k^{3+2b}$, approaching $k^2$ as $\mathrm{w}\to1$. Radiation domination instead gives the familiar cubic leading power, with logarithmic corrections. A sufficiently narrow scalar peak can also produce an intermediate $k^{2-2|b|}$ segment before the finite-width regime is reached~\citep{Domenech:2020kqm,Pi:2020otn}. The high-frequency decline provides additional information about the width of the scalar spectrum, helping to distinguish a change in $\Delta$ from a change in the background equation of state.

These statements apply above the reheating scale. For a finite-width source, modes with $k\ll k_{\rm rh}$ recover a radiation-era cubic leading power after matching across reheating~\citep{Domenech:2020kqm}. The kernel used here assumes $k\gg k_{\rm rh}$ and does not perform that matching~\citep{Witkowski:2022mtg}. We use $k\gtrsim10k_{\rm rh}$, or $f\gtrsim2.4\times10^{-3}\,{\rm Hz}$, as a conservative assumption for its interpretation. We nevertheless note that a joint measurement of the peak, the tails, and a reheating break can help constrain the scalar width and the duration and equation of state of the pre-radiation epoch.

We also find that with the fixed reheating ratio with unit abundance, $f_{\rm pbh}=1$, there are cases that violate the extra-radiation constraint. When choosing $\Delta N_{\rm eff}^{\rm GW}=(5.6\times10^{-6})^{-1}\int\Omega_{{\rm GW},0}h^2\,\mathrm{d}\ln k$~\citep{Clarke:2020bil}, where $\Delta N_{\rm eff}^{\rm GW}$ expresses the GW energy density as an extra effective number of neutrino species, the asymptotic estimates at $\mathrm{w}=1$ are approximately $0.08$, $0.23$, $0.52$, and $2.19$ for $\Delta=0.1$, $0.5$, $1$, and $2$. For comparison, the CMB bound with adiabatic initial conditions in Ref.~\citep{Clarke:2020bil} corresponds to $\Delta N_{\rm eff}^{\rm GW}\lesssim0.3$. Thus, under these assumptions, the broadest stiff cases are not viable all-dark-matter scenarios.

%
%-----------------------
%-------------------------

%------------------------------------------
%------------------------------------------
%--------------     6. SIGW     
%------------------------------------------
%------------------------------------------
\section{Conclusion}
\label{sec:conclusion}
In this work, we have studied the formation of primordial black holes from a wide curvature power spectrum, allowing collapse to occur in a non-standard epoch with an equation of state $\rm w \geq 1/3$. Most earlier studies assume a sharp, nearly monochromatic spectrum, which simplifies computation but is hard to obtain from a realistic inflationary model. We have therefore worked with a log-normal spectrum of finite width $\Delta$, which is a more natural description of the enhancement produced by a transient ultra-slow-roll phase, and we have treated both $\Delta$ and $\rm w$ as free parameters rather than fixing them. {The collapse threshold was determined by NR simulation of a self-gravitating perfect fluid for different values of the spectrum width $\Delta$ and equation of state $\rm w$. Then the abundance was computed within the peak theory using these thresholds. For comparison, we also compute the threshold using two calibrations of the semi-analytical $q$-function prescription. The results show a relative difference within $1\text{--}12\%$. } 

%For the collapse threshold, we first used the two prescriptions of the $q$-function method, which tracks the shape of the collapsing profile and is claimed to work for a general equation of state $\rm w \geq 1/3$. However, it was found that the resulting threshold from this method differs from the NR value. The differences grows as the power spectrum gets wider and also with increasing equation of state $\rm w$. Therefore, we used NR to determine the threshold for all values of $\Delta$ and $\rm w$.

The threshold turned out to be sensitive to both parameters. For a fixed width, the threshold $\mu_{\rm th}$ grows as $\rm w$ increases. The width acts in the same direction, so a broader spectrum raises $\mu_{\rm th}$, and a larger perturbation amplitude is needed before a peak can collapse. {Comparing the NR threshold with the $q$-function prescriptions, it was found that the best agreement is for narrow power spectra and around the radiation domination case for the $wq$-fit prescription with $\epsilon_\mu^{\rm fit} = 1\text{--}3\%$ and around $\rm w = 1/2$ for the $wq$-generic prescription with $\epsilon_\mu^{\rm generic} = 2.8\text{--}3.7\%$. In general, the relative differences grow for the stiffer case and for wider power spectra so that $\epsilon_\mu$ reaches about $12\%$ for the $wq$-fit prescription and $8\%$ for the $wq$-generic prescription. Because the PBH abundance is exponentially sensitive to the threshold, we use the NR threshold for all cases of $\Delta$ and $\rm w$. } 
%Comparing the NR threshold with the $q$-function method, we found an agreement within $2.8\%$-$7.8\%$, in which the best agreement is for narrow spectrum and ${\rm w = 1/2}$. The relative difference rises as the spectrum gets wider and as the equation of state away form ${\rm w = 1/2}$. Since the PBH computation is sensitive to the threshold, we did a full NR analysis, and determine the threshold for all chosen values of $\Delta$ and $\rm w$.}
%The $q$-function method shows an interesting limit for narrow spectra. When $\Delta = 0.1$, this method gives a threshold only up to $\rm w \approx 0.6$, because beyond that point the maximum of the compaction function can no longer reach $\delta_c$. This does not mean that PBHs cannot form. By doing a full NR analysis, we determine the threshold for all chosen values of $\Delta$ and $\rm w$.

{These threshold differences carry over to the abundance calculation. For the fixed reheating ratio $k_{\rm rh}/k_p=0.01$, the integrated scalar amplitude $A$ required for $f_{\rm PBH}^{\rm tot}=1$ increases with both $\Delta$ and $\mathrm{w}$. The curvature-spectrum peak height, however, decreases as the spectrum broadens. Holding $A$ at the value that gives unit abundance in radiation domination instead reveals a strong suppression of the PBH fraction as $\mathrm{w}$ increases, together with a shift of the mass-function peak toward smaller $M/M_{k_p}$.}

{The related SIGW spectra retain information about both the primordial spectrum and the expansion history. A broader scalar spectrum produces a wider GW peak and a more extended high-frequency tail, while a stiffer background enhances the signal and changes its infrared slope above reheating. Combining these features is more informative than the peak amplitude alone, although the low-frequency turnover requires an explicit treatment of the transition to radiation domination. The extra-radiation constraint also excludes some of the broadest stiff examples at the adopted reheating scale, when PBHs are assumed to constitute all of the dark matter.}

Taken together, our results show that the width of the power spectrum and the equation of state at horizon re-entry are not minor details but control parameters of PBH formation. It is understood that the width of the power spectrum and the equation of state of the non-standard epoch change the threshold, the mass function, the abundance, and the gravitational wave signal. These dependencies are lost in the standard treatment, where the power spectrum is monochromatic and $\rm w$ is fixed at $1/3$ to model a radiation-dominant phase. A natural next step is to connect these parameters to specific inflationary and post-inflationary models so that a measured spectral width or a reconstructed equation of state can be turned into a statement about the early thermal history of the universe. 
%------------------------------------------
%------------------------------------------
%------------------------------------------
%------------------------------------------
\begin{acknowledgments}
The authors would like to thank Albert Escriv\`a, Shi Pi, Misao Sasaki, and Chul-Moon Yoo for several stimulating discussions. Y. is supported by the Science and Engineering Research Board (SERB), DST, Government of India, under the Grant Agreement number CRG/2022/004120 (Core Research Grant).
C.J. is supported by NSFC Grants No. E414660101, and No. 12547104, and by the Fundamental Research Funds for the Central Universities under Grants No. E4EQ6604X2 and No. E3ER6601A2. Y.Z. is supported by the Fundamental Research Funds for the Central Universities, and by NSFC Grant No. 12475060, Project 24ZR1472400 sponsored by Natural Science Foundation of Shanghai, and Shanghai
Pujiang Program 24PJA134.

\end{acknowledgments}

%%%%%%%%%%%%%%%%%%%%%%%%%%%%%%%%%%%%%%%%%%%%%%%%%%%%%%%%%%%%%
%%%%%%%%%%%%%%%%%%    Appendix  %%%%%%%%%%%%%%%%%%%%%%%%%%%%%

\appendix
\section{The BSSN formalism of numerical relativity}
\label{app:NR}

We evolve the spacetime with the Baumgarte--Shapiro--Shibata--Nakamura
(BSSN) formulation in spherical symmetry~\citep{Baumgarte_1998,PhysRevD.52.5428,Alcubierre:2011pkc}.
The spacetime line element in a $3+1$ decomposition is
\begin{equation}
    ds^2=-\alpha^2dt^2+\gamma_{ij}\left(dx^i+\beta^i dt\right)
    \left(dx^j+\beta^jdt\right),
    \label{app:spacetime_metric}
\end{equation}
where $\gamma_{ij}$ is the spatial metric, $\alpha$ is the lapse, and $\beta^i$
is the shift. We decompose the spatial metric and extrinsic curvature as
\begin{equation}
    \gamma_{ij}=e^{4\chi}\widetilde\gamma_{ij},
    \qquad
    K_{ij}=e^{4\chi}\left(\widetilde A_{ij}
    +\frac{1}{3}\widetilde\gamma_{ij}K\right).
    \label{app:conformal_decomposition}
\end{equation}
The conformal connection functions are evolved as independent variables. In
the reference-metric formulation they are
$\widehat\Delta^i\equiv\widetilde\gamma^{jk}
(\widetilde\Gamma^i{}_{jk}-\widehat\Gamma^i{}_{jk})$, where
$\widehat\Gamma^i{}_{jk}$ is the connection of the flat reference metric in
spherical coordinates.

For spherical symmetry, the spatial line element becomes
\begin{equation}
    dl^2=e^{4\chi(r,t)}\left[a(r,t)\,dr^2
    +r^2b(r,t)\,d\Omega^2\right],
    \label{app:spherical_metric}
\end{equation}
where $a=\widetilde\gamma_{rr}$ and
$b=\widetilde\gamma_{\theta\theta}/r^2$. We use the mixed components
$A_a=\widetilde A^r{}_r$ and
$A_b=\widetilde A^\theta{}_\theta=\widetilde A^\varphi{}_\varphi$.
Tracelessness gives $A_a+2A_b=0$.

The spherical BSSN evolution equations are
\begin{align}
    \partial_t\chi={}&\beta^r\partial_r\chi-\frac{1}{6}\alpha K,
    \label{app:evolution_chi}\\
    \partial_t a={}&\beta^r\partial_r a+2a\partial_r\beta^r
    -2\alpha aA_a,
    \label{app:evolution_a}\\
    \partial_t b={}&\beta^r\partial_r b+2b\frac{\beta^r}{r}
    -2\alpha bA_b,
    \label{app:evolution_b}\\
    \partial_t K={}&\beta^r\partial_rK-D^2\alpha
    +\alpha\left(A_a^2+2A_b^2+\frac{1}{3}K^2\right)
    +4\pi\alpha\left(\rho+S_a+2S_b\right),
    \label{app:evolution_K}\\
    \partial_t A_a={}&\beta^r\partial_rA_a
    -\left(D^rD_r\alpha-\frac{1}{3}D^2\alpha\right)
    +\alpha\left(R^r{}_r-\frac{1}{3}R\right)
    +\alpha KA_a-\frac{16\pi}{3}\alpha(S_a-S_b).
    \label{app:evolution_Aa}
\end{align}
The radial conformal connection evolves according to
\begin{align}
    \partial_t\widehat\Delta^r={}&
    \beta^r\partial_r\widehat\Delta^r
    -\widehat\Delta^r\partial_r\beta^r
    +\frac{1}{a}\partial_r^2\beta^r
    +\frac{2}{b}\partial_r\left(\frac{\beta^r}{r}\right)
    \notag\\
    &-\frac{2}{a}\left(A_a\partial_r\alpha
    +\alpha\partial_rA_a\right)
    +2\alpha\left[A_a\widehat\Delta^r
    -\frac{2}{rb}(A_a-A_b)\right]
    \notag\\
    &+\frac{2\alpha}{a}\left[
    \partial_rA_a-\frac{2}{3}\partial_rK
    +6A_a\partial_r\chi
    +(A_a-A_b)\left(\frac{2}{r}
    +\frac{\partial_rb}{b}\right)-8\pi S_r\right].
    \label{app:evolution_Delta}
\end{align}
Here $D_i$ is the covariant derivative associated with $\gamma_{ij}$,
$R^i{}_j$ and $R$ are the spatial Ricci tensor and scalar, and the matter
projections are $\rho=n_\mu n_\nu T^{\mu\nu}$,
$S_i=-\gamma_{i\mu}n_\nu T^{\mu\nu}$,
$S_a=S^r{}_r$, and $S_b=S^\theta{}_\theta=S^\varphi{}_\varphi$.

The derivatives of the lapse that enter
Eqs.~\eqref{app:evolution_K} and~\eqref{app:evolution_Aa} are
\begin{align}
    D^2\alpha={}&\frac{e^{-4\chi}}{a}\left[
    \partial_r^2\alpha-\partial_r\alpha\left(
    \frac{\partial_ra}{2a}-\frac{\partial_rb}{b}
    -2\partial_r\chi-\frac{2}{r}\right)\right],
    \label{app:lapse_laplacian}\\
    D^rD_r\alpha={}&\frac{e^{-4\chi}}{a}\left[
    \partial_r^2\alpha-\partial_r\alpha\left(
    \frac{\partial_ra}{2a}+2\partial_r\chi\right)\right].
    \label{app:lapse_rr}
\end{align}
The nontrivial Ricci component and the Ricci scalar can be written as
\begin{align}
    R^r{}_r={}&-\frac{e^{-4\chi}}{a}\Bigg[
    \frac{\partial_r^2a}{2a}-a\partial_r\widehat\Delta^r
    -\frac{3}{4}\left(\frac{\partial_ra}{a}\right)^2
    +\frac{1}{2}\left(\frac{\partial_rb}{b}\right)^2
    -\frac{1}{2}\widehat\Delta^r\partial_ra
    +\frac{\partial_ra}{rb}
    \notag\\
    &\quad+\frac{2}{r^2}\left(1-\frac{a}{b}\right)
    \left(1+\frac{r\partial_rb}{b}\right)
    +4\partial_r^2\chi
    -2\partial_r\chi\left(
    \frac{\partial_ra}{a}-\frac{\partial_rb}{b}-\frac{2}{r}\right)
    \Bigg],
    \label{app:ricci_rr}\\
    R={}&-\frac{e^{-4\chi}}{a}\Bigg[
    \frac{\partial_r^2a}{2a}+\frac{\partial_r^2b}{b}
    -a\partial_r\widehat\Delta^r
    -\left(\frac{\partial_ra}{a}\right)^2
    +\frac{1}{2}\left(\frac{\partial_rb}{b}\right)^2
    +\frac{2}{rb}\left(3-\frac{a}{b}\right)\partial_rb
    \notag\\
    &\quad+\frac{4}{r^2}\left(1-\frac{a}{b}\right)
    +8\left[\partial_r^2\chi+(\partial_r\chi)^2\right]
    -8\partial_r\chi\left(
    \frac{\partial_ra}{2a}-\frac{\partial_rb}{b}-\frac{2}{r}\right)
    \Bigg].
    \label{app:ricci_scalar}
\end{align}

\subsection{General-relativistic hydrodynamics}

The matter source is a perfect fluid with
\begin{equation}
    T_{\mu\nu}=(\rho_{\rm fl}+P)u_\mu u_\nu+Pg_{\mu\nu},
    \qquad P= {\rm w} \rho_{\rm fl},
    \label{app:perfect_fluid}
\end{equation}
where $\rho_{\rm fl}$ is the fluid energy density in its rest frame and
$u^\mu$ is the four-velocity. The equation-of-state parameter $\rm w$ is constant
in each simulation. For an Eulerian observer, the energy and momentum
densities are
\begin{equation}
    \rho=(\rho_{\rm fl}+P)W^2-P,
    \qquad
    S_i=(\rho_{\rm fl}+P)W^2v_i,
    \label{app:eulerian_sources}
\end{equation}
where $W=(1-v^2)^{-1/2}$ and $v^i$ is the Eulerian three-velocity.

The fluid equations follow from baryon-number and energy-momentum
conservation~\citep{10.1093/acprof:oso/9780199205677.001.0001}
\begin{align}
    \nabla_\mu(\rho_0u^\mu)
    &\equiv\frac{1}{\sqrt{-g}}\partial_\mu
    \left(\sqrt{-g}\,\rho_0u^\mu\right)=0,
    \label{app:baryon_conservation}\\
    \nabla_\mu T^\mu{}_{\nu}
    &\equiv\frac{1}{\sqrt{-g}}\partial_\mu
    \left(\sqrt{-g}\,T^\mu{}_{\nu}\right)
    -\Gamma^\lambda{}_{\mu\nu}T^\mu{}_{\lambda}=0.
    \label{app:stress_conservation}
\end{align}
We introduce the conserved variables
\begin{align}
    D&\equiv\rho_0W,
    \label{app:conserved_D}\\
    S_i&\equiv(\rho_{\rm fl}+P)W^2v_i,
    \label{app:conserved_S}\\
    \mathcal E&\equiv(\rho_{\rm fl}+P)W^2-P-D.
    \label{app:conserved_E}
\end{align}
The rest-mass component $D$ is retained in the general system and is zero for
the positive-pressure collapse simulations considered here.
In terms of the BSSN variables, the evolution equations take the form
\begin{align}
    (\partial_t-\mathcal L_\beta)D={}&
    -D_k(\alpha Dv^k)+\alpha KD,
    \label{app:fluid_D}\\
    (\partial_t-\mathcal L_\beta)S_i={}&
    -D_k\left[\alpha\left(S_iv^k+\delta_i{}^kP\right)\right]
    -(\mathcal E+D)D_i\alpha+\alpha KS_i,
    \label{app:fluid_S}\\
    (\partial_t-\mathcal L_\beta)\mathcal E={}&
    (\mathcal E+D+P)\left(\alpha v^mv^nK_{mn}
    -v^m\partial_m\alpha\right)
    \notag\\
    &-D_k\left[\alpha v^k(\mathcal E+P)\right]
    +\alpha K(\mathcal E+P).
    \label{app:fluid_E}
\end{align}
For the barotropic equation of state in Eq.~\eqref{app:perfect_fluid}, the
primitive variables are recovered analytically from a quadratic equation,
avoiding an iterative root search~\citep{Staelens:2019sza}.

\subsection{Gauge conditions}

We evolve the lapse according to a cosmologically adapted Bona--Mass\'o condition,
\begin{equation}
\partial_t\alpha=-\mu_L\alpha^p\bigl(K_{\rm phys}-\langle K_{\rm phys}\rangle\bigr),
\end{equation}
where $K_{\rm phys}=\widehat K+2\Theta$.
We use $0.1\leq\mu_L\leq1$ and $1\leq p\leq2$, choosing values which maintain stable horizon-penetrating slices. In most cases, a constant zero shift is already sufficient for many cases to gauaranty stability until the appearence of the apparent horizon.  However, for some cases, we found useful to evolve the shift with the cosmologically scaled
Gamma-driver of Ref.~\citep{Ning:2026jkk}. In our spherical
reference-metric notation, this reads
\begin{equation}
\begin{aligned}
    \partial_t\beta^r &= b_0\left[\frac{a_{\rm ini}}{a_{\rm bg}(t)}\right]^2 B^r,\\
    \partial_t B^r &= \partial_t\widehat\Delta^r
    -\eta_{{\rm GD},0}\frac{a_{\rm ini}}{a_{\rm bg}(t)}B^r,
\end{aligned}
\label{app:cosmological_shift}
\end{equation}
where $B^r$ is an auxiliary shift variable, $a_{\rm bg}(t)$ is the background
scale factor, $a_{\rm ini}=a_{\rm bg}(t_{\rm ini})$, and $b_0$ and
$\eta_{{\rm GD},0}$ are positive constants.

\subsection{Mass and apparent horizons}

Apparent horizons are located using the outgoing null expansion, while the
Misner--Sharp compactness provides an independent diagnostic. For the
spherically symmetric metric in Eq.~\eqref{app:spherical_metric}, the outgoing and
ingoing null expansions are
\begin{equation}
    \Theta_{\pm}=\pm\frac{e^{-2\chi}}{\sqrt{a}}
    \left(4\,\partial_r\chi+\frac{2}{r}
    +\frac{\partial_rb}{b}\right)
    +A_a-\frac{2}{3}K.
    \label{app:null_expansions}
\end{equation}
The areal radius of a coordinate sphere is
$R=re^{2\chi}\sqrt{b}$. The corresponding Misner--Sharp mass is
\begin{equation}
    M_{\rm MS}=\frac{R}{2}\left[1+
    \left(\frac{\partial_tR-\beta^r\partial_rR}{\alpha}\right)^2
    -\frac{e^{-4\chi}}{a}(\partial_rR)^2\right],
    \label{app:misner_sharp_mass}
\end{equation}
where
\begin{align}
    \partial_tR={}&e^{2\chi}r\sqrt{b}
    \left(2\,\partial_t\chi+\frac{1}{2}
    \frac{\partial_tb}{b}\right),\\
    \partial_rR={}&e^{2\chi}\sqrt{b}
    \left(1+2r\,\partial_r\chi+\frac{r}{2}
    \frac{\partial_rb}{b}\right).
\end{align}
We identify the apparent horizon with the outermost root of
$\Theta_+=0$ for which $\Theta_-<0$. The negative ingoing expansion
distinguishes the black-hole horizon from a cosmological turnaround surface.
At the apparent horizon, $2M_{\rm MS}/R=1$, and the black-hole mass is
$M_{\rm BH}=M_{\rm MS}=R_{\rm AH}/2$.

\subsection{Constraint monitoring}

We monitor the Hamiltonian and radial momentum constraints throughout each
evolution. In the spherical BSSN limit they are
% We also monitor the conformal-connection constraint.
\begin{align}
    \mathcal H={}&R-(A_a^2+2A_b^2)+\frac{2}{3}K^2-16\pi\rho,
    \label{app:hamiltonian_constraint}\\
    \mathcal M_r={}&\partial_rA_a-\frac{2}{3}\partial_rK
    +6A_a\partial_r\chi
    +(A_a-A_b)\left(\frac{2}{r}+\frac{\partial_rb}{b}\right)
    -8\pi S_r,
    \label{app:momentum_constraint}
%    \\
%    \mathcal G^r={}&\widehat\Delta^r-\frac{1}{a}\left[
%    \frac{\partial_ra}{2a}-\frac{\partial_rb}{b}
%    -\frac{2}{r}\left(1-\frac{a}{b}\right)\right].
%    \label{app:connection_constraint}
\end{align}
It is convenient to evaluate the relative constraint violation, which for the
Hamiltonian constraint reads
\begin{equation}
    \mathcal H_{\rm rel}\equiv
    \frac{R-(A_a^2+2A_b^2)+\tfrac{2}{3}K^2-16\pi\rho}
    {|R|+|A_a^2+2A_b^2|+\tfrac{2}{3}|K^2|+16\pi|\rho|}.
    \label{app:relative_hamiltonian_constraint}
\end{equation}
We require $|\mathcal H_{\rm rel}|<10^{-2}$ throughout the physical domain
outside the apparent horizon.
The initial density is obtained from $\mathcal H=0$, while
$\mathcal H$, $\mathcal M_r$,
% $\mathcal G^r$,
the algebraic relation
$A_a+2A_b=0$, and the conformal-determinant condition $ab^2=1$ are checked
during the evolution.

\begin{figure}[!t]
    \centering
    \includegraphics[width=0.92\columnwidth]
    {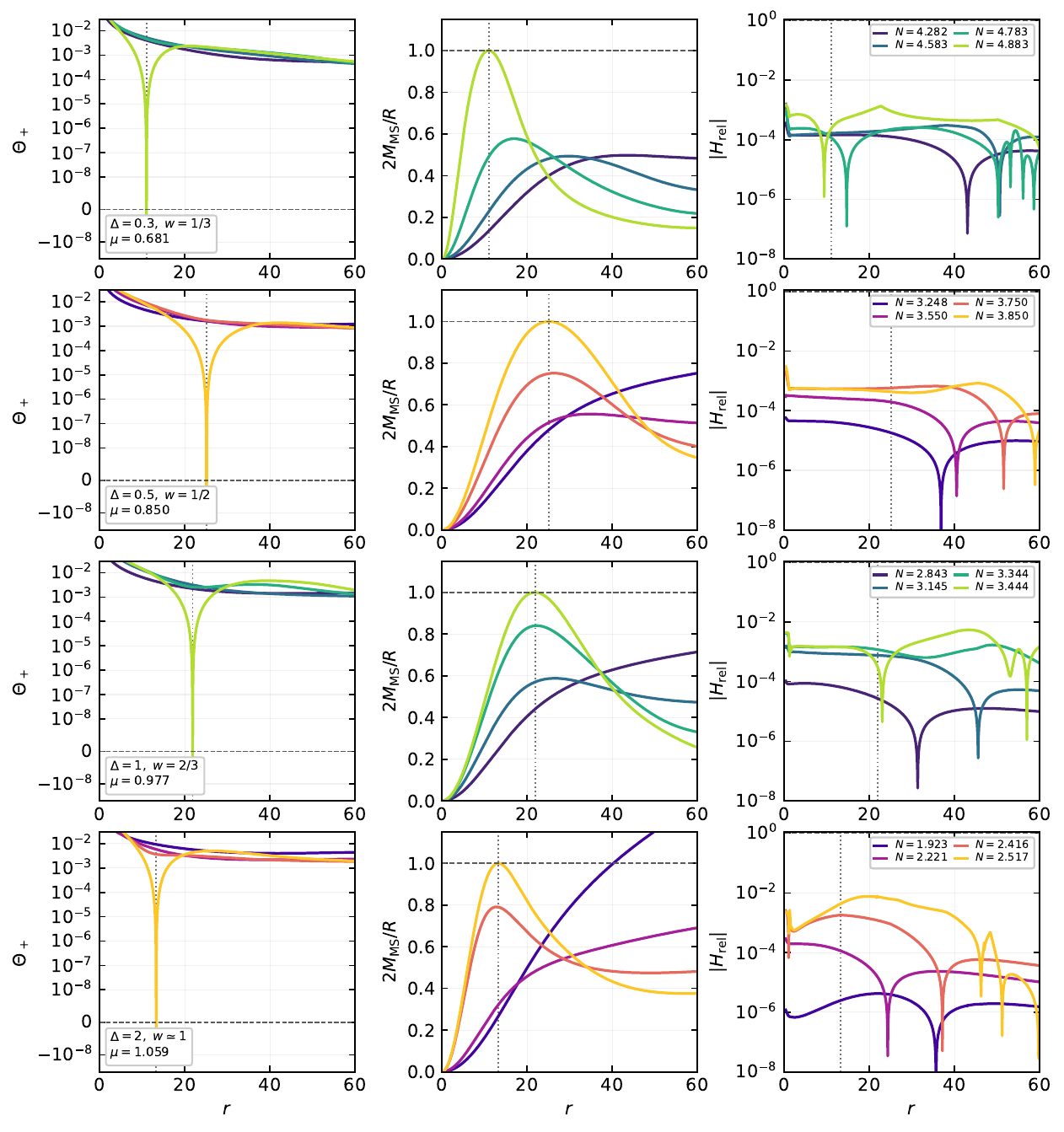}
    \caption{Near-threshold apparent-horizon diagnostics for PBH-forming simulations. From top to bottom, the rows correspond to
    $(\Delta,\rm w,\mu)=(0.3,1/3,0.681)$, $(0.5,1/2,0.850)$,
    $(1,2/3,0.977)$, and $(2,1,1.059)$. The columns show the null
    expansion $\Theta_+$, the Misner--Sharp compactness $2M_{\rm MS}/R$,
    and the relative Hamiltonian-constraint violation
    $|\mathcal H_{\rm rel}|$. Each row contains four e-fold times $N$, with
    the latest profile marking the first checkpoint containing an apparent
    horizon. The vertical dotted line indicates $r_{\rm AH}$, while the
    horizontal dashed lines in the left and central columns mark
    $\Theta_+=0$ and $2M_{\rm MS}/R=1$, respectively.}
    \label{fig:NR_AH_diagnostics}
\end{figure}

The four examples in Fig.~\ref{fig:NR_AH_diagnostics} show that the zero of
$\Theta_+$ and unit Misner--Sharp compactness coincide at the marked apparent horizon radius, while $|\mathcal H_{\rm rel}|$ remains below $10^{-2}$.

\clearpage

%%%%%%%%%%%%%%%%%%%%%%%%%%%%%%%%%%%%%%%%%%%%%%%%%%%%%%%%%%%%%%%%
%%%%%%%%%%%%%%%%%%    References   %%%%%%%%%%%%%%%%%%%%%%%%%%%%%

\bibliographystyle{apsrev4-1}
\bibliography{pbh.bib}

@article{Zeldovich:1967lct,
    author = "Zel'dovich, Ya. B. and Novikov, I. D.",
    title = "{The Hypothesis of Cores Retarded during Expansion and the Hot Cosmological Model}",
    journal = "Sov. Astron.",
    volume = "10",
    pages = "602",
    year = "1967"
}

@article{Hawking:1971ei,
    author = "Hawking, Stephen",
    title = "{Gravitationally collapsed objects of very low mass}",
    journal = "Mon. Not. Roy. Astron. Soc.",
    volume = "152",
    pages = "75",
    year = "1971"
}

@article{Laha:2019ssq,
    author = "Laha, Ranjan",
    title = "{Primordial Black Holes as a Dark Matter Candidate Are Severely Constrained by the Galactic Center 511 keV $\gamma$ -Ray Line}",
    eprint = "1906.09994",
    archivePrefix = "arXiv",
    primaryClass = "astro-ph.HE",
    reportNumber = "CERN-TH-2019-099",
    doi = "10.1103/PhysRevLett.123.251101",
    journal = "Phys. Rev. Lett.",
    volume = "123",
    number = "25",
    pages = "251101",
    year = "2019"
}

@article{Carr:1974nx,
    author = "Carr, Bernard J. and Hawking, S.W.",
    title = "{Black holes in the early Universe}",
    journal = "Mon. Not. Roy. Astron. Soc.",
    volume = "168",
    pages = "399--415",
    year = "1974"
}

@article{Carr:1975qj,
    author = "Carr, Bernard J.",
    title = "{The Primordial black hole mass spectrum}",
    doi = "10.1086/153853",
    journal = "Astrophys. J.",
    volume = "201",
    pages = "1--19",
    year = "1975"
}

@article{Young:2019yug,
    author = "Young, Sam and Musco, Ilia and Byrnes, Christian T.",
    title = "{Primordial black hole formation and abundance: contribution from the non-linear relation between the density and curvature perturbation}",
    eprint = "1904.00984",
    archivePrefix = "arXiv",
    primaryClass = "astro-ph.CO",
    doi = "10.1088/1475-7516/2019/11/012",
    journal = "JCAP",
    volume = "11",
    pages = "012",
    year = "2019"
}

@article{LIGOScientific:2016aoc,
    author = "Abbott, B. P. and others",
    collaboration = "LIGO Scientific, Virgo",
    title = "{Observation of Gravitational Waves from a Binary Black Hole Merger}",
    eprint = "1602.03837",
    archivePrefix = "arXiv",
    primaryClass = "gr-qc",
    reportNumber = "LIGO-P150914",
    doi = "10.1103/PhysRevLett.116.061102",
    journal = "Phys. Rev. Lett.",
    volume = "116",
    number = "6",
    pages = "061102",
    year = "2016"
}

@article{LIGOScientific:2016dsl,
    author = "Abbott, B. P. and others",
    collaboration = "LIGO Scientific, Virgo",
    title = "{Binary Black Hole Mergers in the first Advanced LIGO Observing Run}",
   
    reportNumber = "LIGO-P1600088",
    
    journal = "Phys. Rev. X",
    volume = "6",
    number = "4",
    pages = "041015",
    year = "2016",
    note = "[Erratum: Phys.Rev.X 8, 039903 (2018)]"
}

@article{LIGOScientific:2016sjg,
    author = "Abbott, B. P. and others",
    collaboration = "LIGO Scientific, Virgo",
    title = "{GW151226: Observation of Gravitational Waves from a 22-Solar-Mass Binary Black Hole Coalescence}",
    eprint = "1606.04855",
    archivePrefix = "arXiv",
    primaryClass = "gr-qc",
    reportNumber = "LIGO-P151226",
    doi = "10.1103/PhysRevLett.116.241103",
    journal = "Phys. Rev. Lett.",
    volume = "116",
    number = "24",
    pages = "241103",
    year = "2016"
}

@article{Escriva:2020tak,
    author = "Escriv\`a, Albert and Germani, Cristiano and Sheth, Ravi K.",
    title = "{Analytical thresholds for black hole formation in general cosmological backgrounds}",
    eprint = "2007.05564",
    archivePrefix = "arXiv",
    primaryClass = "gr-qc",
    reportNumber = "ICCUB-20-016",
    doi = "10.1088/1475-7516/2021/01/030",
    journal = "JCAP",
    volume = "01",
    pages = "030",
    year = "2021"
}

@article{Escriva:2019phb,
    author = "Escriv\`a, Albert and Germani, Cristiano and Sheth, Ravi K.",
    title = "{Universal threshold for primordial black hole formation}",
    eprint = "1907.13311",
    archivePrefix = "arXiv",
    primaryClass = "gr-qc",
    reportNumber = "ICC-19-013",
    doi = "10.1103/PhysRevD.101.044022",
    journal = "Phys. Rev. D",
    volume = "101",
    number = "4",
    pages = "044022",
    year = "2020"
}

@article{LIGOScientific:2016wyt,
    author = "Abbott, Benjamin P. and others",
    collaboration = "LIGO Scientific, Virgo",
    title = "{The basic physics of the binary black hole merger GW150914}",
    eprint = "1608.01940",
    archivePrefix = "arXiv",
    primaryClass = "gr-qc",
    doi = "10.1002/andp.201600209",
    journal = "Annalen Phys.",
    volume = "529",
    number = "1-2",
    pages = "1600209",
    year = "2017"
}

@article{LIGOScientific:2017bnn,
    author = "Abbott, Benjamin P. and others",
    collaboration = "LIGO Scientific, VIRGO",
    title = "{GW170104: Observation of a 50-Solar-Mass Binary Black Hole Coalescence at Redshift 0.2}",
    eprint = "1706.01812",
    archivePrefix = "arXiv",
    primaryClass = "gr-qc",
    reportNumber = "LIGO-P170104",
    doi = "10.1103/PhysRevLett.118.221101",
    journal = "Phys. Rev. Lett.",
    volume = "118",
    number = "22",
    pages = "221101",
    year = "2017",
    note = "[Erratum: Phys.Rev.Lett. 121, 129901 (2018)]"
}

@article{LIGOScientific:2017vox,
    author = "Abbott, B. P. and others",
    collaboration = "LIGO Scientific, Virgo",
    title = "{GW170608: Observation of a 19-solar-mass Binary Black Hole Coalescence}",
    eprint = "1711.05578",
    archivePrefix = "arXiv",
    primaryClass = "astro-ph.HE",
    reportNumber = "LIGO-DOCUMENT-P170608-V8",
    doi = "10.3847/2041-8213/aa9f0c",
    journal = "Astrophys. J. Lett.",
    volume = "851",
    pages = "L35",
    year = "2017"
}

@article{LIGOScientific:2017ycc,
    author = "Abbott, B. P. and others",
    collaboration = "LIGO Scientific, Virgo",
    title = "{GW170814: A Three-Detector Observation of Gravitational Waves from a Binary Black Hole Coalescence}",
    eprint = "1709.09660",
    archivePrefix = "arXiv",
    primaryClass = "gr-qc",
    doi = "10.1103/PhysRevLett.119.141101",
    journal = "Phys. Rev. Lett.",
    volume = "119",
    number = "14",
    pages = "141101",
    year = "2017"
}

@article{Fernandez:2019kyb,
    author = "Fernandez, Nicolas and Profumo, Stefano",
    title = "{Unraveling the origin of black holes from effective spin measurements with LIGO-Virgo}",
    eprint = "1905.13019",
    archivePrefix = "arXiv",
    primaryClass = "astro-ph.HE",
    reportNumber = "SCIPP 19/02",
    doi = "10.1088/1475-7516/2019/08/022",
    journal = "JCAP",
    volume = "08",
    pages = "022",
    year = "2019"
}

@article{Carr:2020gox,
    author = "Carr, Bernard and Kohri, Kazunori and Sendouda, Yuuiti and Yokoyama, Jun'ichi",
    title = "{Constraints on primordial black holes}",
    eprint = "2002.12778",
    archivePrefix = "arXiv",
    primaryClass = "astro-ph.CO",
    reportNumber = "RESCEU-03/20; KEK-Cosmo-249; KEK-TH-2199; IPMU20-0024",
    doi = "10.1088/1361-6633/ac1e31",
    journal = "Rept. Prog. Phys.",
    volume = "84",
    number = "11",
    pages = "116902",
    year = "2021"
}

@article{Carr:2021bzv,
    author = "Carr, Bernard and Kuhnel, Florian",
    title = "{Primordial black holes as dark matter candidates}",
    eprint = "2110.02821",
    archivePrefix = "arXiv",
    primaryClass = "astro-ph.CO",
    doi = "10.21468/SciPostPhysLectNotes.48",
    journal = "SciPost Phys. Lect. Notes",
    volume = "48",
    pages = "1",
    year = "2022"
}

@article{PhysRevD.50.7173,
  title = {Inflation and primordial black holes as dark matter},
  author = {Ivanov, P. and Naselsky, P. and Novikov, I.},
  journal = {Phys. Rev. D},
  volume = {50},
  issue = {12},
  pages = {7173--7178},
  numpages = {0},
  year = {1994},
  month = {Dec},
  publisher = {American Physical Society},
  doi = {10.1103/PhysRevD.50.7173},
  url = {https://link.aps.org/doi/10.1103/PhysRevD.50.7173}
}

@article{Yokoyama:1998pt,
    author = "Yokoyama, Jun'ichi",
    title = "{Chaotic new inflation and formation of primordial black holes}",
    eprint = "astro-ph/9802357",
    archivePrefix = "arXiv",
    reportNumber = "YITP-98-10, SU-ITP-98-04",
    doi = "10.1103/PhysRevD.58.083510",
    journal = "Phys. Rev. D",
    volume = "58",
    pages = "083510",
    year = 1998
}

@article{Garcia-Bellido:2017mdw,
    author = "Garcia-Bellido, Juan and Ruiz Morales, Ester",
    title = "{Primordial black holes from single field models of inflation}",
    eprint = "1702.03901",
    archivePrefix = "arXiv",
    primaryClass = "astro-ph.CO",
    reportNumber = "IFT-UAM-CSIC-17-007, CERN-TH-2017-196",
    doi = "10.1016/j.dark.2017.09.007",
    journal = "Phys. Dark Univ.",
    volume = "18",
    pages = "47--54",
    year = "2017"
}

@article{Ballesteros:2017fsr,
    author = "Ballesteros, Guillermo and Taoso, Marco",
    title = "{Primordial black hole dark matter from single field inflation}",
    eprint = "1709.05565",
    archivePrefix = "arXiv",
    primaryClass = "hep-ph",
    doi = "10.1103/PhysRevD.97.023501",
    journal = "Phys. Rev. D",
    volume = "97",
    number = "2",
    pages = "023501",
    year = "2018"
}

@article{Kinney:2005vj,
    author = "Kinney, William H.",
    title = "{Horizon crossing and inflation with large eta}",
    eprint = "gr-qc/0503017",
    archivePrefix = "arXiv",
    doi = "10.1103/PhysRevD.72.023515",
    journal = "Phys. Rev. D",
    volume = "72",
    pages = "023515",
    year = "2005"
}

@article{Germani:2017bcs,
    author = "Germani, Cristiano and Prokopec, Tomislav",
    title = "{On primordial black holes from an inflection point}",
    eprint = "1706.04226",
    archivePrefix = "arXiv",
    primaryClass = "astro-ph.CO",
    reportNumber = "ICCUB-17-012",
    doi = "10.1016/j.dark.2017.09.001",
    journal = "Phys. Dark Univ.",
    volume = "18",
    pages = "6--10",
    year = "2017"
}

@article{Biagetti:2018pjj,
    author = "Biagetti, Matteo and Franciolini, Gabriele and Kehagias, Alex and Riotto, Antonio",
    title = "{Primordial Black Holes from Inflation and Quantum Diffusion}",
    eprint = "1804.07124",
    archivePrefix = "arXiv",
    primaryClass = "astro-ph.CO",
    doi = "10.1088/1475-7516/2018/07/032",
    journal = "JCAP",
    volume = "07",
    pages = "032",
    year = "2018"
}

@article{Kawasaki:1997ju,
    author = "Kawasaki, M. and Sugiyama, N. and Yanagida, T.",
    title = "{Primordial black hole formation in a double inflation model in supergravity}",
    eprint = "hep-ph/9710259",
    archivePrefix = "arXiv",
    reportNumber = "KUNS-1470, ICRR-413-98-9",
    doi = "10.1103/PhysRevD.57.6050",
    journal = "Phys. Rev. D",
    volume = "57",
    pages = "6050--6056",
    year = "1998"
}

@article{Crawford:1982yz,
    author = "Crawford, Matt and Schramm, David N.",
    title = "{Spontaneous Generation of Density Perturbations in the Early Universe}",
    reportNumber = "EFI-82-04-CHICAGO",
    doi = "10.1038/298538a0",
    journal = "Nature",
    volume = "298",
    pages = "538--540",
    year = "1982"
}

@article{Harada:2024jxl,
    author = "Harada, Tomohiro",
    title = "{Primordial Black Holes: Formation, Spin and Type II}",
    eprint = "2409.01934",
    archivePrefix = "arXiv",
    primaryClass = "gr-qc",
    reportNumber = "RUP-24-16",
    doi = "10.3390/universe10120444",
    journal = "Universe",
    volume = "10",
    number = "12",
    pages = "444",
    year = "2024"
}

@article{HosseiniMansoori:2023mqh,
    author = "Hosseini Mansoori, Seyed Ali and Felegray, Fereshteh and Talebian, Alireza and Sami, Mohammad",
    title = "{PBHs and GWs from {\ensuremath{\mathbb{T}}}$^{2}$-inflation and NANOGrav 15-year data}",
    eprint = "2307.06757",
    archivePrefix = "arXiv",
    primaryClass = "astro-ph.CO",
    doi = "10.1088/1475-7516/2023/08/067",
    journal = "JCAP",
    volume = "08",
    pages = "067",
    year = "2023"
}

@article{Pi:2024ert,
    author = "Pi, Shi and Sasaki, Misao and Takhistov, Volodymyr and Wang, Jianing",
    title = "{Primordial Black Hole formation from power spectrum with finite-width}",
    eprint = "2501.00295",
    archivePrefix = "arXiv",
    primaryClass = "astro-ph.CO",
    reportNumber = "YITP-24-184, KEK-QUP-2024-0028, KEK-TH-2676, KEK-Cosmo-0369",
    doi = "10.1088/1475-7516/2025/09/045",
    journal = "JCAP",
    volume = "09",
    pages = "045",
    year = "2025"
}

@article{Inui:2024fgk,
    author = "Inui, Ryoto and Joana, Cristian and Motohashi, Hayato and Pi, Shi and Tada, Yuichiro and Yokoyama, Shuichiro",
    title = "{Primordial black holes and induced gravitational waves from logarithmic non-Gaussianity}",
    eprint = "2411.07647",
    archivePrefix = "arXiv",
    primaryClass = "astro-ph.CO",
    doi = "10.1088/1475-7516/2025/03/021",
    journal = "JCAP",
    volume = "03",
    pages = "021",
    year = "2025"
}

@article{Hawking:1982ga,
    author = "Hawking, S. W. and Moss, I. G. and Stewart, J. M.",
    title = "{Bubble Collisions in the Very Early Universe}",
    reportNumber = "Print-82-0180 (CAMBRIDGE)",
    doi = "10.1103/PhysRevD.26.2681",
    journal = "Phys. Rev. D",
    volume = "26",
    pages = "2681",
    year = "1982"
}

@article{Blanco-Pillado:2017rnf,
    author = "Blanco-Pillado, Jose J. and Olum, Ken D. and Siemens, Xavier",
    title = "{New limits on cosmic strings from gravitational wave observation}",
    eprint = "1709.02434",
    archivePrefix = "arXiv",
    primaryClass = "astro-ph.CO",
    doi = "10.1016/j.physletb.2018.01.050",
    journal = "Phys. Lett. B",
    volume = "778",
    pages = "392--396",
    year = "2018"
}

@article{HAWKING1989237,
title = {Black holes from cosmic strings},
journal = {Physics Letters B},
volume = {231},
number = {3},
pages = {237-239},
year = {1989},
issn = {0370-2693},
doi = {https://doi.org/10.1016/0370-2693(89)90206-2},
author = {S.W. Hawking},
}

@article{Rubin:2001yw,
    author = "Rubin, Sergei G. and Sakharov, Alexander S. and Khlopov, Maxim Yu.",
    title = "{The Formation of primary galactic nuclei during phase transitions in the early universe}",
    eprint = "hep-ph/0106187",
    archivePrefix = "arXiv",
    doi = "10.1134/1.1385631",
    journal = "J. Exp. Theor. Phys.",
    volume = "91",
    pages = "921--929",
    year = "2001"
}

@article{Garriga:2015fdk,
    author = "Garriga, Jaume and Vilenkin, Alexander and Zhang, Jun",
    title = "{Black holes and the multiverse}",
    eprint = "1512.01819",
    archivePrefix = "arXiv",
    primaryClass = "hep-th",
    doi = "10.1088/1475-7516/2016/02/064",
    journal = "JCAP",
    volume = "02",
    pages = "064",
    year = "2016"
}

@article{Mishra:2019pzq,
    author = "Mishra, Swagat S. and Sahni, Varun",
    title = "{Primordial Black Holes from a tiny bump/dip in the Inflaton potential}",
    eprint = "1911.00057",
    archivePrefix = "arXiv",
    primaryClass = "gr-qc",
    doi = "10.1088/1475-7516/2020/04/007",
    journal = "JCAP",
    volume = "04",
    pages = "007",
    year = "2020"
}

@article{Ragavendra:2020sop,
    author = "Ragavendra, H. V. and Saha, Pankaj and Sriramkumar, L. and Silk, Joseph",
    title = "{Primordial black holes and secondary gravitational waves from ultraslow roll and punctuated inflation}",
    eprint = "2008.12202",
    archivePrefix = "arXiv",
    primaryClass = "astro-ph.CO",
    doi = "10.1103/PhysRevD.103.083510",
    journal = "Phys. Rev. D",
    volume = "103",
    number = "8",
    pages = "083510",
    year = "2021"
}

@article{Bhattacharya:2019bvk,
    author = "Bhattacharya, Sukannya and Mohanty, Subhendra and Parashari, Priyank",
    title = "{Primordial black holes and gravitational waves in nonstandard cosmologies}",
    eprint = "1912.01653",
    archivePrefix = "arXiv",
    primaryClass = "astro-ph.CO",
    doi = "10.1103/PhysRevD.102.043522",
    journal = "Phys. Rev. D",
    volume = "102",
    number = "4",
    pages = "043522",
    year = "2020"
}

@article{Bhattacharya:2020lhc,
    author = "Bhattacharya, Sukannya and Mohanty, Subhendra and Parashari, Priyank",
    title = "{Implications of the NANOGrav result on primordial gravitational waves in nonstandard cosmologies}",
    eprint = "2010.05071",
    archivePrefix = "arXiv",
    primaryClass = "astro-ph.CO",
    doi = "10.1103/PhysRevD.103.063532",
    journal = "Phys. Rev. D",
    volume = "103",
    number = "6",
    pages = "063532",
    year = "2021"
}

@article{Bhattacharya:2023ztw,
    author = "Bhattacharya, Sukannya",
    title = "{Primordial Black Hole Formation in Non-Standard Post-Inflationary Epochs}",
    eprint = "2302.12690",
    archivePrefix = "arXiv",
    primaryClass = "astro-ph.CO",
    doi = "10.3390/galaxies11010035",
    journal = "Galaxies",
    volume = "11",
    number = "1",
    pages = "35",
    year = "2023"
}

@article{Yogesh:2025hll,
    author = "Yogesh and Mohammadi, Abolhassan",
    title = "{Nonstandard Thermal History and Formation of Primordial Black Holes and SIGWs in Einstein{\textendash}Gauss{\textendash}Bonnet Gravity}",
    eprint = "2501.01867",
    archivePrefix = "arXiv",
    primaryClass = "gr-qc",
    doi = "10.3847/1538-4357/adcee5",
    journal = "Astrophys. J.",
    volume = "986",
    number = "1",
    pages = "35",
    year = "2025"
}

@article{Solbi:2024zhl,
    author = "Solbi, Milad and Karami, Kayoomars",
    title = "{Primordial black holes in non-minimal Gauss\textendash{}Bonnet inflation in light of the PTA data}",
    eprint = "2403.00021",
    archivePrefix = "arXiv",
    primaryClass = "gr-qc",
    doi = "10.1140/epjc/s10052-024-13271-x",
    journal = "Eur. Phys. J. C",
    volume = "84",
    number = "9",
    pages = "918",
    year = "2024"
}

@article{Ashrafzadeh:2023ndt,
    author = "Ashrafzadeh, Ali and Karami, Kayoomars",
    title = "{Primordial Black Holes in Scalar Field Inflation Coupled to the Gauss\textendash{}Bonnet Term with Fractional Power-law Potentials}",
    eprint = "2309.16356",
    archivePrefix = "arXiv",
    primaryClass = "astro-ph.CO",
    doi = "10.3847/1538-4357/ad293f",
    journal = "Astrophys. J.",
    volume = "965",
    number = "1",
    pages = "11",
    year = "2024"
}

@article{Bardeen:1985tr,
    author = "Bardeen, James M. and Bond, J. R. and Kaiser, Nick and Szalay, A. S.",
    title = "{The Statistics of Peaks of Gaussian Random Fields}",
    reportNumber = "FERMILAB-PUB-85-148-A, NSF-ITP-85-93",
    doi = "10.1086/164143",
    journal = "Astrophys. J.",
    volume = "304",
    pages = "15--61",
    year = "1986"
}

@article{Niemeyer:1997mt,
    author = "Niemeyer, Jens C. and Jedamzik, K.",
    title = "{Near-critical gravitational collapse and the initial mass function of primordial black holes}",
    eprint = "astro-ph/9709072",
    archivePrefix = "arXiv",
    doi = "10.1103/PhysRevLett.80.5481",
    journal = "Phys. Rev. Lett.",
    volume = "80",
    pages = "5481--5484",
    year = "1998"
}

@article{Planck:2018jri,
    author = "Akrami, Y. and others",
    collaboration = "Planck",
    title = "{Planck 2018 results. X. Constraints on inflation}",
    eprint = "1807.06211",
    archivePrefix = "arXiv",
    primaryClass = "astro-ph.CO",
    doi = "10.1051/0004-6361/201833887",
    journal = "Astron. Astrophys.",
    volume = "641",
    pages = "A10",
    year = "2020"
}

@article{Braglia:2020eai,
    author = "Braglia, Matteo and Hazra, Dhiraj Kumar and Finelli, Fabio and Smoot, George F. and Sriramkumar, L. and Starobinsky, Alexei A.",
    title = "{Generating PBHs and small-scale GWs in two-field models of inflation}",
    eprint = "2005.02895",
    archivePrefix = "arXiv",
    primaryClass = "astro-ph.CO",
    doi = "10.1088/1475-7516/2020/08/001",
    journal = "JCAP",
    volume = "08",
    pages = "001",
    year = "2020"
}

@article{Chen:2024roo,
    author = {Chen, Chao and Dimopoulos, Konstantinos and Er\"oncel, Cem and Ghoshal, Anish},
    title = "{Enhanced primordial gravitational waves from a stiff postinflationary era due to an oscillating inflaton}",
    eprint = "2405.01679",
    archivePrefix = "arXiv",
    primaryClass = "hep-ph",
    doi = "10.1103/PhysRevD.110.063554",
    journal = "Phys. Rev. D",
    volume = "110",
    number = "6",
    pages = "063554",
    year = "2024"
}

@article{Musco:2018rwt,
    author = "Musco, Ilia",
    title = "{Threshold for primordial black holes: Dependence on the shape of the cosmological perturbations}",
    eprint = "1809.02127",
    archivePrefix = "arXiv",
    primaryClass = "gr-qc",
    doi = "10.1103/PhysRevD.100.123524",
    journal = "Phys. Rev. D",
    volume = "100",
    number = "12",
    pages = "123524",
    year = "2019"
}

@article{Pi:2022zxs,
    author = "Pi, Shi and Wang, Jianing",
    title = "{Primordial black hole formation in Starobinsky's linear potential model}",
    eprint = "2209.14183",
    archivePrefix = "arXiv",
    primaryClass = "astro-ph.CO",
    reportNumber = "IPMU22-0047",
    doi = "10.1088/1475-7516/2023/06/018",
    journal = "JCAP",
    volume = "06",
    pages = "018",
    year = "2023"
}

@article{Cacciapaglia:2025xqd,
    author = "Cacciapaglia, Giacomo and Cheong, Dhong Yeon and Deandrea, Aldo and Isnard, Wanda and Park, Seong Chan and Wang, Xinpeng and Zhang, Ying-li",
    title = "{Composite Hybrid Inflation : Primordial Black Holes and Stochastic Gravitational Waves}",
    eprint = "2506.06655",
    archivePrefix = "arXiv",
    primaryClass = "hep-ph",
    month = "6",
    year = "2025"
}

@article{Wang:2025lti,
    author = "Wang, Xinpeng and Sasaki, Misao and Zhang, Ying-li",
    title = "{Dual primordial black hole formation scenario}",
    eprint = "2505.09337",
    archivePrefix = "arXiv",
    primaryClass = "astro-ph.CO",
    reportNumber = "YITP-25-67",
    doi = "10.1103/ct3g-6d9k",
    journal = "Phys. Rev. D",
    volume = "113",
    number = "12",
    pages = "L121304",
    year = "2026"
}

@article{Kim:2025dyi,
    author = "Kim, Jinsu and Wang, Xinpeng and Zhang, Ying-li and Ren, Zhongzhou",
    title = "{Enhancement of primordial curvature perturbations in R $^{3}$-corrected Starobinsky-Higgs inflation}",
    eprint = "2504.12035",
    archivePrefix = "arXiv",
    primaryClass = "astro-ph.CO",
    doi = "10.1088/1475-7516/2025/09/011",
    journal = "JCAP",
    volume = "09",
    pages = "011",
    year = "2025"
}

@article{Bird:2016dcv,
    author = {Bird, Simeon and Cholis, Ilias and Mu\~noz, Julian B. and Ali-Haïmoud, Yacine and Kamionkowski, Marc and Kovetz, Ely D. and Raccanelli, Alvise and Riess, Adam G.},
    title = "{Did LIGO detect dark matter?}",
    eprint = "1603.00464",
    archivePrefix = "arXiv",
    primaryClass = "astro-ph.CO",
    doi = "10.1103/PhysRevLett.116.201301",
    journal = "Phys. Rev. Lett.",
    volume = "116",
    number = "20",
    pages = "201301",
    year = "2016"
}

@article{Sasaki:2016jop,
    author = "Sasaki, Misao and Suyama, Teruaki and Tanaka, Takahiro and Yokoyama, Shuichiro",
    title = "{Primordial Black Hole Scenario for the Gravitational-Wave Event GW150914}",
    eprint = "1603.08338",
    archivePrefix = "arXiv",
    primaryClass = "astro-ph.CO",
    reportNumber = "RESCEU-17-16, RUP-16-7, YITP-16-43",
    doi = "10.1103/PhysRevLett.117.061101",
    journal = "Phys. Rev. Lett.",
    volume = "117",
    number = "6",
    pages = "061101",
    year = "2016",
    note = "[Erratum: Phys.Rev.Lett. 121, 059901 (2018)]"
}

@article{Clesse:2016vqa,
    author = "Clesse, Sebastien and Garc\'\i{}a-Bellido, Juan",
    title = "{The clustering of massive Primordial Black Holes as Dark Matter: measuring their mass distribution with Advanced LIGO}",
    eprint = "1603.05234",
    archivePrefix = "arXiv",
    primaryClass = "astro-ph.CO",
    reportNumber = "TTK-16-10, IFT-UAM-CSIC-16-027",
    doi = "10.1016/j.dark.2016.10.002",
    journal = "Phys. Dark Univ.",
    volume = "15",
    pages = "142--147",
    year = "2017"
}

@article{DeLuca:2020qqa,
    author = "De Luca, V. and Franciolini, G. and Pani, P. and Riotto, A.",
    title = "{Primordial Black Holes Confront LIGO/Virgo data: Current situation}",
    eprint = "2005.05641",
    archivePrefix = "arXiv",
    primaryClass = "astro-ph.CO",
    doi = "10.1088/1475-7516/2020/06/044",
    journal = "JCAP",
    volume = "06",
    pages = "044",
    year = "2020"
}

@article{Coughlan:1983ci,
    author = "Coughlan, G. D. and Fischler, W. and Kolb, Edward W. and Raby, S. and Ross, Graham G.",
    title = "{Cosmological Problems for the Polonyi Potential}",
    reportNumber = "LA-UR-83-1423",
    doi = "10.1016/0370-2693(83)91091-2",
    journal = "Phys. Lett. B",
    volume = "131",
    pages = "59--64",
    year = "1983"
}

@article{Starobinsky:1994bd,
    author = "Starobinsky, Alexei A. and Yokoyama, Junichi",
    title = "{Equilibrium state of a selfinteracting scalar field in the De Sitter background}",
    eprint = "astro-ph/9407016",
    archivePrefix = "arXiv",
    reportNumber = "YITP-U-94-12",
    doi = "10.1103/PhysRevD.50.6357",
    journal = "Phys. Rev. D",
    volume = "50",
    pages = "6357--6368",
    year = "1994"
}

@article{Peebles:1998qn,
    author = "Peebles, P. J. E. and Vilenkin, A.",
    title = "{Quintessential inflation}",
    eprint = "astro-ph/9810509",
    archivePrefix = "arXiv",
    doi = "10.1103/PhysRevD.59.063505",
    journal = "Phys. Rev. D",
    volume = "59",
    pages = "063505",
    year = "1999"
}

@article{Ahmad:2019jbm,
    author = "Ahmad, Safia and De Felice, Antonio and Jaman, Nur and Kuroyanagi, Sachiko and Sami, M.",
    title = "{Baryogenesis in the paradigm of quintessential inflation}",
    eprint = "1908.03742",
    archivePrefix = "arXiv",
    primaryClass = "gr-qc",
    reportNumber = "YITP-19-133",
    doi = "10.1103/PhysRevD.100.103525",
    journal = "Phys. Rev. D",
    volume = "100",
    number = "10",
    pages = "103525",
    year = "2019"
}

@article{Allahverdi:2020bys,
    author = "Allahverdi, Rouzbeh and others",
    title = "{The First Three Seconds: a Review of Possible Expansion Histories of the Early Universe}",
    eprint = "2006.16182",
    archivePrefix = "arXiv",
    primaryClass = "astro-ph.CO",
    reportNumber = "FERMILAB-PUB-20-242-A, KCL-PH-TH/2020-33, KEK-Cosmo-257,
  KEK-TH-2231, IPMU20-0070, PI/UAN-2020-674FT, RUP-20-22",
    month = jun,
    journal = "arXiv preprint",
    year = "2020"
}

@article{Kawai:2021edk,
    author = "Kawai, Shinsuke and Kim, Jinsu",
    title = "{Primordial black holes from Gauss-Bonnet-corrected single field inflation}",
    eprint = "2108.01340",
    archivePrefix = "arXiv",
    primaryClass = "astro-ph.CO",
    reportNumber = "CERN-TH-2021-115",
    doi = "10.1103/PhysRevD.104.083545",
    journal = "Phys. Rev. D",
    volume = "104",
    number = "8",
    pages = "083545",
    year = "2021"
}

@article{Joana:2026myf,
    author = "Joana, Cristian",
    title = "{Primordial black holes forming during kination: the trapped, the overdense, and the void}",
    eprint = "2607.20423",
    archivePrefix = "arXiv",
    primaryClass = "astro-ph.CO",
    month = "7",
    year = "2026"
}

@article{Joana:2025gqf,
    author = "Joana, Cristian and Yuwen, Zi-Yan",
    title = "{Primordial black holes from primordial voids}",
    eprint = "2510.11611",
    archivePrefix = "arXiv",
    primaryClass = "astro-ph.CO",
    doi = "10.1103/j3hw-d5cx",
    journal = "Phys. Rev. D",
    volume = "113",
    number = "2",
    pages = "023518",
    year = "2026"
}

@article{Germani:2025hcu,
    author = "Germani, Cristiano and Montell{\`a}, Laia",
    title = "{Trichotomy of primordial black holes initial conditions}",
    eprint = "2510.02006",
    archivePrefix = "arXiv",
    primaryClass = "gr-qc",
    doi = "10.1103/6ysb-nbt8",
    journal = "Phys. Rev. D",
    volume = "113",
    number = "6",
    pages = "064054",
    year = "2026"
}

@article{Deng:2017uwc,
    author = "Deng, Heling and Vilenkin, Alexander",
    title = "{Primordial black hole formation by vacuum bubbles}",
    eprint = "1710.02865",
    archivePrefix = "arXiv",
    primaryClass = "gr-qc",
    doi = "10.1088/1475-7516/2017/12/044",
    journal = "JCAP",
    volume = "12",
    pages = "044",
    year = "2017"
}

@article{Yuwen:2026hxu,
    author = "Yuwen, Zi-Yan and Joana, Cristian and Wang, Shao-Jiang and Cai, Rong-Gen",
    title = "{Primordial black hole formation in bulk-viscous cosmology}",
    eprint = "2606.26532",
    archivePrefix = "arXiv",
    primaryClass = "gr-qc",
    month = "6",
    year = "2026"
}

@article{Baumgarte_1998,
    author = "Baumgarte, Thomas W. and Shapiro, Stuart L.",
    title = "{Numerical integration of Einstein's field equations}",
    doi = "10.1103/PhysRevD.59.024007",
    journal = "Phys. Rev. D",
    volume = "59",
    number = "2",
    pages = "024007",
    year = "1999"
}

@article{PhysRevD.52.5428,
    author = "Shibata, Masaru and Nakamura, Takashi",
    title = "{Evolution of three-dimensional gravitational waves: Harmonic slicing case}",
    doi = "10.1103/PhysRevD.52.5428",
    journal = "Phys. Rev. D",
    volume = "52",
    number = "10",
    pages = "5428--5444",
    year = "1995"
}

@article{Alcubierre:2011pkc,
    author = "Alcubierre, Miguel and Mendez, Martha D.",
    title = "{Formulations of the 3+1 evolution equations in curvilinear coordinates}",
    eprint = "1010.4013",
    archivePrefix = "arXiv",
    primaryClass = "gr-qc",
    doi = "10.1007/s10714-011-1202-x",
    journal = "Gen. Rel. Grav.",
    volume = "43",
    pages = "2769--2806",
    year = "2011"
}

@book{10.1093/acprof:oso/9780199205677.001.0001,
    author = "Alcubierre, Miguel",
    title = "{Introduction to 3+1 Numerical Relativity}",
    publisher = "Oxford University Press",
    isbn = "9780199205677",
    doi = "10.1093/acprof:oso/9780199205677.001.0001",
    year = "2008"
}

@article{Staelens:2019sza,
    author = {Staelens, Fran\c{c}ois and Rekier, J\'er\'emy and F{\"u}zfa, Andr\'e},
    title = "{Universality of spherical collapse with respect to the matter type: The case of a barotropic fluid with linear equation of state}",
    eprint = "1912.00677",
    archivePrefix = "arXiv",
    primaryClass = "gr-qc",
    doi = "10.1007/s10714-021-02804-4",
    journal = "Gen. Rel. Grav.",
    volume = "53",
    number = "4",
    pages = "38",
    year = "2021"
}

@article{Yoo:2018kvb,
    author = "Yoo, Chul-Moon and Harada, Tomohiro and Garriga, Jaume and Kohri, Kazunori",
    title = "{Primordial black hole abundance from random Gaussian curvature perturbations and a local density threshold}",
    eprint = "1805.03946",
    archivePrefix = "arXiv",
    primaryClass = "astro-ph.CO",
    reportNumber = "RUP-18-15, KEK-Cosmo-225, KEK-TH-2052",
    doi = "10.1093/ptep/pty120",
    journal = "PTEP",
    volume = "2018",
    number = "12",
    pages = "123E01",
    year = "2018",
    note = "[Erratum: PTEP 2024, 049202 (2024)]"
}

@article{Atal:2019erb,
    author = "Atal, Vicente and Cid, Judith and Escriv{\`a}, Albert and Garriga, Jaume",
    title = "{PBH in single field inflation: the effect of shape dispersion and non-Gaussianities}",
    eprint = "1908.11357",
    archivePrefix = "arXiv",
    primaryClass = "astro-ph.CO",
    doi = "10.1088/1475-7516/2020/05/022",
    journal = "JCAP",
    volume = "05",
    pages = "022",
    year = "2020"
}

@article{Germani:2019zez,
    author = "Germani, Cristiano and Sheth, Ravi K.",
    title = "{Nonlinear statistics of primordial black holes from Gaussian curvature perturbations}",
    eprint = "1912.07072",
    archivePrefix = "arXiv",
    primaryClass = "astro-ph.CO",
    reportNumber = "ICCUB-19-021",
    doi = "10.1103/PhysRevD.101.063520",
    journal = "Phys. Rev. D",
    volume = "101",
    number = "6",
    pages = "063520",
    year = "2020"
}

@article{Yoo:2020dkz,
    author = "Yoo, Chul-Moon and Harada, Tomohiro and Hirano, Shin'ichi and Kohri, Kazunori",
    title = "{Abundance of Primordial Black Holes in Peak Theory for an Arbitrary Power Spectrum}",
    eprint = "2008.02425",
    archivePrefix = "arXiv",
    primaryClass = "astro-ph.CO",
    reportNumber = "RUP-20-25, KEK-Cosmo-261, KEK-TH-2245",
    doi = "10.1093/ptep/ptaa155",
    journal = "PTEP",
    volume = "2021",
    number = "1",
    pages = "013E02",
    year = "2021",
    note = "[Erratum: PTEP 2024, 049203 (2024)]"
}

@article{Kitajima:2021fpq,
    author = "Kitajima, Naoya and Tada, Yuichiro and Yokoyama, Shuichiro and Yoo, Chul-Moon",
    title = "{Primordial black holes in peak theory with a non-Gaussian tail}",
    eprint = "2109.00791",
    archivePrefix = "arXiv",
    primaryClass = "astro-ph.CO",
    reportNumber = "TU-1130",
    doi = "10.1088/1475-7516/2021/10/053",
    journal = "JCAP",
    volume = "10",
    pages = "053",
    year = "2021"
}

@ARTICLE{1974ApJ...187..425P,
       author = {{Press}, William H. and {Schechter}, Paul},
        title = "{Formation of Galaxies and Clusters of Galaxies by Self-Similar Gravitational Condensation}",
      journal = {\apj},
         year = 1974,
        month = feb,
       volume = {187},
        pages = {425-438},
          doi = {10.1086/152650},
       adsurl = {https://ui.adsabs.harvard.edu/abs/1974ApJ...187..425P}
}

@article{Young:2020xmk,
    author = "Young, Sam and Musso, Marcello",
    title = "{Application of peaks theory to the abundance of primordial black holes}",
    eprint = "2001.06469",
    archivePrefix = "arXiv",
    primaryClass = "astro-ph.CO",
    doi = "10.1088/1475-7516/2020/11/022",
    journal = "JCAP",
    volume = "11",
    pages = "022",
    year = "2020"
}

@article{ACT:2025fju,
    author = "Louis, Thibaut and others",
    collaboration = "Atacama Cosmology Telescope",
    title = "{The Atacama Cosmology Telescope: DR6 power spectra, likelihoods and {\ensuremath{\Lambda}}CDM parameters}",
    eprint = "2503.14452",
    archivePrefix = "arXiv",
    primaryClass = "astro-ph.CO",
    reportNumber = "FERMILAB-PUB-25-0071-PPD",
    doi = "10.1088/1475-7516/2025/11/062",
    journal = "JCAP",
    volume = "11",
    pages = "062",
    year = "2025"
}

@article{Pi:2024jwt,
    author = "Pi, Shi",
    title = "{Non-Gaussianities in primordial black hole formation and induced gravitational waves}",
    eprint = "2404.06151",
    archivePrefix = "arXiv",
    primaryClass = "astro-ph.CO",
    month = "4",
    year = "2024"
}

@article{LISACosmologyWorkingGroup:2022jok,
    author = "Auclair, Pierre and others",
    collaboration = "LISA Cosmology Working Group",
    title = "{Cosmology with the Laser Interferometer Space Antenna}",
    eprint = "2204.05434",
    archivePrefix = "arXiv",
    primaryClass = "astro-ph.CO",
    reportNumber = "LISA CosWG-22-03, FERMILAB-PUB-22-349-SCD",
    doi = "10.1007/s41114-023-00045-2",
    journal = "Living Rev. Rel.",
    volume = "26",
    number = "1",
    pages = "5",
    year = "2023"
}

@article{Domenech:2024rks,
    author = "Dom\`enech, Guillem and Pi, Shi and Wang, Ao and Wang, Jianing",
    title = "{Induced gravitational wave interpretation of PTA data: a complete study for general equation of state}",
    eprint = "2402.18965",
    archivePrefix = "arXiv",
    primaryClass = "astro-ph.CO",
    doi = "10.1088/1475-7516/2024/08/054",
    journal = "JCAP",
    volume = "08",
    pages = "054",
    year = "2024"
}

@article{Liu:2023hpw,
    author = "Liu, Lang and Wu, You and Chen, Zu-Cheng",
    title = "{Simultaneously probing the sound speed and equation of state of the early Universe with pulsar timing arrays}",
    eprint = "2310.16500",
    archivePrefix = "arXiv",
    primaryClass = "astro-ph.CO",
    doi = "10.1088/1475-7516/2024/04/011",
    journal = "JCAP",
    volume = "04",
    pages = "011",
    year = "2024"
}

@article{NANOGrav:2023gor,
    author = "Agazie, Gabriella and others",
    collaboration = "NANOGrav",
    title = "{The NANOGrav 15 yr Data Set: Evidence for a Gravitational-wave Background}",
    eprint = "2306.16213",
    archivePrefix = "arXiv",
    primaryClass = "astro-ph.HE",
    doi = "10.3847/2041-8213/acdac6",
    journal = "Astrophys. J. Lett.",
    volume = "951",
    number = "1",
    pages = "L8",
    year = "2023"
}

@article{Mohammadi:2025avz,
    author = "Mohammadi, Abolhassan and Yogesh and Wu, Qiang and Zhu, Tao",
    title = "{Spinning Primordial Black Holes and Scalar Induced Gravitational Waves from Single Field Inflation}",
    eprint = "2512.05435",
    archivePrefix = "arXiv",
    primaryClass = "astro-ph.CO",
    doi = "10.3847/1538-4357/ae47ed",
    journal = "Astrophys. J.",
    volume = "1000",
    number = "1",
    pages = "101",
    year = "2026"
}

@article{Caprini:2018mtu,
	author = "Caprini, Chiara and Figueroa, Daniel G.",
	title = "{Cosmological Backgrounds of Gravitational Waves}",
	eprint = "1801.04268",
	archivePrefix = "arXiv",
	primaryClass = "astro-ph.CO",
	doi = "10.1088/1361-6382/aac608",
	journal = "Class. Quant. Grav.",
	volume = "35",
	number = "16",
	pages = "163001",
	year = "2018"
}

@article{Christensen:2018iqi,
	author = "Christensen, Nelson",
	title = "{Stochastic Gravitational Wave Backgrounds}",
	eprint = "1811.08797",
	archivePrefix = "arXiv",
	primaryClass = "gr-qc",
	doi = "10.1088/1361-6633/aae6b5",
	journal = "Rept. Prog. Phys.",
	volume = "82",
	number = "1",
	pages = "016903",
	year = "2019"
}

@article{Figueroa:2019paj,
	author = "Figueroa, Daniel G. and Tanin, Erwin H.",
	title = "{Ability of LIGO and LISA to probe the equation of state of the early Universe}",
	eprint = "1905.11960",
	archivePrefix = "arXiv",
	primaryClass = "astro-ph.CO",
	doi = "10.1088/1475-7516/2019/08/011",
	journal = "JCAP",
	volume = "08",
	pages = "011",
	year = "2019"
}

@article{Bernal:2019lpc,
	author = "Bernal, Nicol\'as and Hajkarim, Fazlollah",
	title = "{Primordial Gravitational Waves in Nonstandard Cosmologies}",
	eprint = "1905.10410",
	archivePrefix = "arXiv",
	primaryClass = "astro-ph.CO",
	doi = "10.1103/PhysRevD.100.063502",
	journal = "Phys. Rev. D",
	volume = "100",
	number = "6",
	pages = "063502",
	year = "2019"
}

@article{Bernal:2020ywq,
	author = "Bernal, Nicol\'as and Ghoshal, Anish and Hajkarim, Fazlollah and Lambiase, Gaetano",
	title = "{Primordial Gravitational Wave Signals in Modified Cosmologies}",
	eprint = "2008.04959",
	archivePrefix = "arXiv",
	primaryClass = "gr-qc",
	doi = "10.1088/1475-7516/2020/11/051",
	journal = "JCAP",
	volume = "11",
	pages = "051",
	year = "2020"
}

@article{Espinosa:2018eve,
    author = "Espinosa, Jos\'e Ram\'on and Racco, Davide and Riotto, Antonio",
    title = "{A Cosmological Signature of the SM Higgs Instability: Gravitational Waves}",
    eprint = "1804.07732",
    archivePrefix = "arXiv",
    primaryClass = "hep-ph",
    doi = "10.1088/1475-7516/2018/09/012",
    journal = "JCAP",
    volume = "09",
    pages = "012",
    year = "2018"
}

@article{Kohri:2018awv,
    author = "Kohri, Kazunori and Terada, Takahiro",
    title = "{Semianalytic calculation of gravitational wave spectrum nonlinearly induced from primordial curvature perturbations}",
    eprint = "1804.08577",
    archivePrefix = "arXiv",
    primaryClass = "gr-qc",
    reportNumber = "KEK-TH-2046, KEK-COSMO-223",
    doi = "10.1103/PhysRevD.97.123532",
    journal = "Phys. Rev. D",
    volume = "97",
    number = "12",
    pages = "123532",
    year = "2018"
}

@article{Ananda:2006af,
	author = "Ananda, Kishore N. and Clarkson, Chris and Wands, David",
	title = "{The Cosmological gravitational wave background from primordial density perturbations}",
	eprint = "gr-qc/0612013",
	archivePrefix = "arXiv",
	doi = "10.1103/PhysRevD.75.123518",
	journal = "Phys. Rev. D",
	volume = "75",
	pages = "123518",
	year = "2007"
}

@article{Baumann:2007zm,
	author = "Baumann, Daniel and Steinhardt, Paul J. and Takahashi, Keitaro and Ichiki, Kiyotomo",
	title = "{Gravitational Wave Spectrum Induced by Primordial Scalar Perturbations}",
	eprint = "hep-th/0703290",
	archivePrefix = "arXiv",
	doi = "10.1103/PhysRevD.76.084019",
	journal = "Phys. Rev. D",
	volume = "76",
	pages = "084019",
	year = "2007"
}

@article{Domenech:2021ztg,
	author = "Dom\`enech, Guillem",
	title = "{Scalar Induced Gravitational Waves Review}",
	eprint = "2109.01398",
	archivePrefix = "arXiv",
	primaryClass = "gr-qc",
	doi = "10.3390/universe7110398",
	journal = "Universe",
	volume = "7",
	number = "11",
	pages = "398",
	year = "2021"
}

@article{Domenech:2020kqm,
	author = "Dom\`enech, Guillem and Pi, Shi and Sasaki, Misao",
	title = "{Induced gravitational waves as a probe of thermal history of the universe}",
	eprint = "2005.12314",
	archivePrefix = "arXiv",
	primaryClass = "gr-qc",
	reportNumber = "YITP-20-70, IPMU20-0053",
	doi = "10.1088/1475-7516/2020/08/017",
	journal = "JCAP",
	volume = "08",
	pages = "017",
	year = "2020"
}

@article{Witkowski:2022mtg,
    author = "Witkowski, Lukas T.",
    title = "{SIGWfast: a python package for the computation of scalar-induced gravitational wave spectra}",
    eprint = "2209.05296",
    archivePrefix = "arXiv",
    primaryClass = "astro-ph.CO",
    month = sep,
    year = "2022"
}

@article{Uehara:2025idq,
    author = "Uehara, Koichiro and Escriv{\`a}, Albert and Harada, Tomohiro and Saito, Daiki and Yoo, Chul-Moon",
    title = "{Primordial black hole formation from a type II perturbation in the absence and presence of pressure}",
    eprint = "2505.00366",
    archivePrefix = "arXiv",
    primaryClass = "gr-qc",
    reportNumber = "RUP-25-12; NU-QG-5; KUNS-3055, RUP-25-12",
    doi = "10.1088/1475-7516/2025/08/042",
    journal = "JCAP",
    volume = "08",
    pages = "042",
    year = "2025"
}

@article{Escriva:2025rja,
    author = "Escriv{\`a}, Albert",
    title = "{Threshold for PBH formation in the type-II region and its analytical estimation}",
    eprint = "2504.05814",
    archivePrefix = "arXiv",
    primaryClass = "astro-ph.CO",
    doi = "10.1103/mq67-bbvj",
    journal = "Phys. Rev. D",
    volume = "112",
    number = "10",
    pages = "103527",
    year = "2025"
}

@article{Escriva:2022pnz,
    author = "Escriv{\`a}, Albert and Tada, Yuichiro and Yokoyama, Shuichiro and Yoo, Chul-Moon",
    title = "{Simulation of primordial black holes with large negative non-Gaussianity}",
    eprint = "2202.01028",
    archivePrefix = "arXiv",
    primaryClass = "astro-ph.CO",
    doi = "10.1088/1475-7516/2022/05/012",
    journal = "JCAP",
    volume = "05",
    number = "05",
    pages = "012",
    year = "2022"
}

@article{Papanikolaou:2022hkg,
    author = "Papanikolaou, Theodoros and Tzerefos, Charalampos and Basilakos, Spyros and Saridakis, Emmanuel N.",
    title = "{No constraints for f(T) gravity from gravitational waves induced from primordial black hole fluctuations}",
    eprint = "2205.06094",
    archivePrefix = "arXiv",
    primaryClass = "gr-qc",
    doi = "10.1140/epjc/s10052-022-11157-4",
    journal = "Eur. Phys. J. C",
    volume = "83",
    number = "1",
    pages = "31",
    year = "2023"
}

@article{Banerjee:2022xft,
    author = "Banerjee, Shreya and Papanikolaou, Theodoros and Saridakis, Emmanuel N.",
    title = "{Constraining F(R) bouncing cosmologies through primordial black holes}",
    eprint = "2206.01150",
    archivePrefix = "arXiv",
    primaryClass = "gr-qc",
    doi = "10.1103/PhysRevD.106.124012",
    journal = "Phys. Rev. D",
    volume = "106",
    number = "12",
    pages = "124012",
    year = "2022"
}

@article{Escriva:2022duf,
    author = "Escriv\`a, Albert and Kuhnel, Florian and Tada, Yuichiro",
    title = "{Primordial Black Holes}",
    eprint = "2211.05767",
    archivePrefix = "arXiv",
    primaryClass = "astro-ph.CO",
    doi = "10.1016/B978-0-32-395636-9.00012-8",
    month = "11",
    year = "2022"
}

@article{EPTA:2023fyk,
	author = "Antoniadis, J. and others",
	collaboration = "EPTA, InPTA:",
	title = "{The second data release from the European Pulsar Timing Array - III. Search for gravitational wave signals}",
	eprint = "2306.16214",
	archivePrefix = "arXiv",
	primaryClass = "astro-ph.HE",
	doi = "10.1051/0004-6361/202346844",
	journal = "Astron. Astrophys.",
	volume = "678",
	pages = "A50",
	year = "2023"
}

@article{Shimada:2024eec,
    author = "Shimada, Masaaki and Escriv{\'a}, Albert and Saito, Daiki and Uehara, Koichiro and Yoo, Chul-Moon",
    title = "{Primordial black hole formation from type II fluctuations with primordial non-Gaussianity}",
    eprint = "2411.07648",
    archivePrefix = "arXiv",
    primaryClass = "gr-qc",
    doi = "10.1088/1475-7516/2025/02/018",
    journal = "JCAP",
    volume = "02",
    pages = "018",
    year = "2025"
}

@article{Shibata:1999zs,
    author = "Shibata, Masaru and Sasaki, Misao",
    title = "{Black hole formation in the Friedmann universe: Formulation and computation in numerical relativity}",
    eprint = "gr-qc/9905064",
    archivePrefix = "arXiv",
    reportNumber = "OU-TAP-93",
    doi = "10.1103/PhysRevD.60.084002",
    journal = "Phys. Rev. D",
    volume = "60",
    pages = "084002",
    year = "1999"
}

@article{Peacock:1990zz,
    author = "Peacock, J. A. and Heavens, A. F.",
    title = "{Alternatives to the Press-Schechter cosmological mass function}",
    journal = "Mon. Not. Roy. Astron. Soc.",
    volume = "243",
    pages = "133--143",
    year = "1990"
}

@article{Pi:2020otn,
    author = "Pi, Shi and Sasaki, Misao",
    title = "{Gravitational Waves Induced by Scalar Perturbations with a Lognormal Peak}",
    eprint = "2005.12306",
    archivePrefix = "arXiv",
    primaryClass = "gr-qc",
    reportNumber = "YITP-20-75, YITP-75, IPMU20-0054",
    doi = "10.1088/1475-7516/2020/09/037",
    journal = "JCAP",
    volume = "09",
    pages = "037",
    year = "2020"
}

@article{Choptuik:1992jv,
    author = "Choptuik, Matthew W.",
    title = "{Universality and scaling in gravitational collapse of a massless scalar field}",
    reportNumber = "FPRINT-92-33",
    doi = "10.1103/PhysRevLett.70.9",
    journal = "Phys. Rev. Lett.",
    volume = "70",
    pages = "9--12",
    year = "1993"
}

@article{Harada:2015yda,
    author = "Harada, Tomohiro and Yoo, Chul-Moon and Nakama, Tomohiro and Koga, Yasutaka",
    title = "{Cosmological long-wavelength solutions and primordial black hole formation}",
    eprint = "1503.03934",
    archivePrefix = "arXiv",
    primaryClass = "gr-qc",
    reportNumber = "RUP-15-5, RESCEU-4-15",
    doi = "10.1103/PhysRevD.91.084057",
    journal = "Phys. Rev. D",
    volume = "91",
    number = "8",
    pages = "084057",
    year = "2015"
}

@article{Hazra:2010ve,
    author = "Hazra, Dhiraj Kumar and Aich, Moumita and Jain, Rajeev Kumar and Sriramkumar, L. and Souradeep, Tarun",
    title = "{Primordial features due to a step in the inflaton potential}",
    eprint = "1005.2175",
    archivePrefix = "arXiv",
    primaryClass = "astro-ph.CO",
    doi = "10.1088/1475-7516/2010/10/008",
    journal = "JCAP",
    volume = "10",
    pages = "008",
    year = "2010"
}

@article{Kawasaki:1999na,
    author = "Kawasaki, M. and Kohri, Kazunori and Sugiyama, Naoshi",
    title = "{Cosmological constraints on late time entropy production}",
    eprint = "astro-ph/9811437",
    archivePrefix = "arXiv",
    reportNumber = "RESCEU-7-99, KUNS-1546",
    doi = "10.1103/PhysRevLett.82.4168",
    journal = "Phys. Rev. Lett.",
    volume = "82",
    pages = "4168",
    year = "1999"
}

@article{Hasegawa:2019jsa,
    author = "Hasegawa, Takuya and Hiroshima, Nagisa and Kohri, Kazunori and Hansen, Rasmus S. L. and Tram, Thomas and Hannestad, Steen",
    title = "{MeV-scale reheating temperature and thermalization of oscillating neutrinos by radiative and hadronic decays of massive particles}",
    eprint = "1908.10189",
    archivePrefix = "arXiv",
    primaryClass = "hep-ph",
    reportNumber = "KEK-TH-2149, KEK-Cosmo-242, RIKEN-iTHEMS-Report-19, IPMU19-0120",
    doi = "10.1088/1475-7516/2019/12/012",
    journal = "JCAP",
    volume = "12",
    pages = "012",
    year = "2019"
}

@article{Carr:1993aq,
    author = "Carr, Bernard J. and Lidsey, James E.",
    title = "{Primordial black holes and generalized constraints on chaotic inflation}",
    reportNumber = "FERMILAB-PUB-93-116-A",
    doi = "10.1103/PhysRevD.48.543",
    journal = "Phys. Rev. D",
    volume = "48",
    pages = "543--553",
    year = "1993"
}

@article{Xu:2023wog,
    author = "Xu, Heng and others",
    title = "{Searching for the Nano-Hertz Stochastic Gravitational Wave Background with the Chinese Pulsar Timing Array Data Release I}",
    eprint = "2306.16216",
    archivePrefix = "arXiv",
    primaryClass = "astro-ph.HE",
    doi = "10.1088/1674-4527/acdfa5",
    journal = "Res. Astron. Astrophys.",
    volume = "23",
    number = "7",
    pages = "075024",
    year = "2023"
}

@ARTICLE{2013PASA...30...17M,
       author = {{Manchester}, R.~N, et. al.},
        title = "{The Parkes Pulsar Timing Array Project}",
      journal = {\pasa},
         year = 2013,
        month = jan,
       volume = {30},
          eid = {e017},
        pages = {e017},
          doi = {10.1017/pasa.2012.017},
archivePrefix = {arXiv},
       eprint = {1210.6130},
 primaryClass = {astro-ph.IM},
       adsurl = {https://ui.adsabs.harvard.edu/abs/2013PASA...30...17M}
}

@article{Hobbs:2013aka,
    author = "Hobbs, G.",
    title = "{The Parkes Pulsar Timing Array}",
    eprint = "1307.2629",
    archivePrefix = "arXiv",
    primaryClass = "astro-ph.IM",
    doi = "10.1088/0264-9381/30/22/224007",
    journal = "Class. Quant. Grav.",
    volume = "30",
    pages = "224007",
    year = "2013"
}

@article{Sato-Polito:2019hws,
    author = "Sato-Polito, Gabriela and Kovetz, Ely D. and Kamionkowski, Marc",
    title = "{Constraints on the primordial curvature power spectrum from primordial black holes}",
    eprint = "1904.10971",
    archivePrefix = "arXiv",
    primaryClass = "astro-ph.CO",
    doi = "10.1103/PhysRevD.100.063521",
    journal = "Phys. Rev. D",
    volume = "100",
    number = "6",
    pages = "063521",
    year = "2019"
}

@article{Gangopadhyay:2026mck,
    author = "Gangopadhyay, Mayukh R.",
    title = "{Primordial Black Hole Formation in Rastall Gravity: Shifted Collapse Threshold and Exponential Abundance Sensitivity}",
    eprint = "2602.19826",
    archivePrefix = "arXiv",
    primaryClass = "astro-ph.CO",
    doi = "10.1016/j.dark.2026.102332",
    journal = "Phys. Dark Univ.",
    volume = "52",
    pages = "102332",
    year = "2026"
}

@article{Choudhury:2024aji,
    author = "Choudhury, Sayantan and Sami, M.",
    title = "{Large fluctuations and primordial black holes}",
    eprint = "2407.17006",
    archivePrefix = "arXiv",
    primaryClass = "gr-qc",
    doi = "10.1016/j.physrep.2024.10.007",
    journal = "Phys. Rept.",
    volume = "1103",
    pages = "1--276",
    year = "2025"
}

@article{Maison:1995cc,
    author = "Maison, Dieter",
    title = "{Nonuniversality of critical behavior in spherically symmetric gravitational collapse}",
    eprint = "gr-qc/9504008",
    archivePrefix = "arXiv",
    reportNumber = "MPI-PHT-95-28",
    doi = "10.1016/0370-2693(95)01381-4",
    journal = "Phys. Lett. B",
    volume = "366",
    pages = "82--84",
    year = "1996"
}

@article{Musco:2012au,
    author = "Musco, Ilia and Miller, John C.",
    title = "{Primordial black hole formation in the early universe: critical behaviour and self-similarity}",
    eprint = "1201.2379",
    archivePrefix = "arXiv",
    primaryClass = "gr-qc",
    doi = "10.1088/0264-9381/30/14/145009",
    journal = "Class. Quant. Grav.",
    volume = "30",
    pages = "145009",
    year = "2013"
}

@article{Escriva:2021pmf,
    author = "Escriv{\`a}, Albert and Romano, Antonio Enea",
    title = "{Effects of the shape of curvature peaks on the size of primordial black holes}",
    eprint = "2103.03867",
    archivePrefix = "arXiv",
    primaryClass = "gr-qc",
    reportNumber = "ICCUB-21-003",
    doi = "10.1088/1475-7516/2021/05/066",
    journal = "JCAP",
    volume = "05",
    pages = "066",
    year = "2021"
}

@article{Solbi:2021wbo,
    author = "Solbi, Milad and Karami, Kayoomars",
    title = "{Primordial black holes and induced gravitational waves in $k$-inflation}",
    eprint = "2102.05651",
    archivePrefix = "arXiv",
    primaryClass = "astro-ph.CO",
    doi = "10.1088/1475-7516/2021/08/056",
    journal = "JCAP",
    volume = "08",
    pages = "056",
    year = "2021"
}

@article{Escriva:2024aeo,
    author = "Escriv{\`a}, Albert and Yoo, Chul-Moon",
    title = "{Nonspherical effects on the mass function of primordial black holes}",
    eprint = "2410.03451",
    archivePrefix = "arXiv",
    primaryClass = "gr-qc",
    doi = "10.1103/4jbp-87wc",
    journal = "Phys. Rev. D",
    volume = "112",
    number = "8",
    pages = "L081304",
    year = "2025"
}

@article{Escriva:2024lmm,
    author = "Escriv{\`a}, Albert and Yoo, Chul-Moon",
    title = "{Simulations of ellipsoidal primordial black hole formation}",
    eprint = "2410.03452",
    archivePrefix = "arXiv",
    primaryClass = "gr-qc",
    doi = "10.1103/PhysRevD.112.083518",
    journal = "Phys. Rev. D",
    volume = "112",
    number = "8",
    pages = "083518",
    year = "2025"
}

@article{Byrnes:2018txb,
    author = "Byrnes, Christian T. and Cole, Philippa S. and Patil, Subodh P.",
    title = "{Steepest growth of the power spectrum and primordial black holes}",
    eprint = "1811.11158",
    archivePrefix = "arXiv",
    primaryClass = "astro-ph.CO",
    doi = "10.1088/1475-7516/2019/06/028",
    journal = "JCAP",
    volume = "06",
    pages = "028",
    year = "2019"
}

@article{Germani:2023ojx,
    author = "Germani, Cristiano and Sheth, Ravi K.",
    title = "{The Statistics of Primordial Black Holes in a Radiation-Dominated Universe: Recent and New Results}",
    eprint = "2308.02971",
    archivePrefix = "arXiv",
    primaryClass = "astro-ph.CO",
    doi = "10.3390/universe9090421",
    journal = "Universe",
    volume = "9",
    number = "9",
    pages = "421",
    year = "2023"
}

@article{Wang:2025hwc,
    author = "Wang, Haonan and Zhang, Ying-li and Suyama, Teruaki",
    title = "{Nearly Monochromatic Primordial Black Holes as total Dark Matter from Bubble Collapse}",
    eprint = "2510.19233",
    archivePrefix = "arXiv",
    primaryClass = "astro-ph.CO",
    month = "10",
    year = "2025"
}

@article{Wang:2024vfv,
    author = "Wang, Xinpeng and Zhang, Ying-li and Sasaki, Misao",
    title = "{Enhanced curvature perturbation and primordial black hole formation in two-stage inflation with a break}",
    eprint = "2404.02492",
    archivePrefix = "arXiv",
    primaryClass = "astro-ph.CO",
    reportNumber = "YITP-24-25",
    doi = "10.1088/1475-7516/2024/07/076",
    journal = "JCAP",
    volume = "07",
    pages = "076",
    year = "2024"
}

@article{Ashrafzadeh:2024oll,
    author = "Ashrafzadeh, A. and Solbi, M. and Heydari, S. and Karami, K.",
    title = "{Primordial black holes in SB SUSY Gauss-Bonnet inflation}",
    eprint = "2407.15445",
    archivePrefix = "arXiv",
    primaryClass = "hep-th",
    doi = "10.1088/1475-7516/2025/01/025",
    journal = "JCAP",
    volume = "01",
    pages = "025",
    year = "2025"
}

@article{Barausse:2020rsu,
    author = "Barausse, Enrico and others",
    title = "{Prospects for Fundamental Physics with LISA}",
    eprint = "2001.09793",
    archivePrefix = "arXiv",
    primaryClass = "gr-qc",
    doi = "10.1007/s10714-020-02691-1",
    journal = "Gen. Rel. Grav.",
    volume = "52",
    number = "8",
    pages = "81",
    year = "2020"
}

@article{Bartolo:2018evs,
    author = "Bartolo, N. and De Luca, V. and Franciolini, G. and Lewis, A. and Peloso, M. and Riotto, A.",
    title = "{Primordial Black Hole Dark Matter: LISA Serendipity}",
    eprint = "1810.12218",
    archivePrefix = "arXiv",
    primaryClass = "astro-ph.CO",
    doi = "10.1103/PhysRevLett.122.211301",
    journal = "Phys. Rev. Lett.",
    volume = "122",
    number = "21",
    pages = "211301",
    year = "2019"
}

@article{Musco:2020jjb,
    author = "Musco, Ilia and De Luca, Valerio and Franciolini, Gabriele and Riotto, Antonio",
    title = "{Threshold for primordial black holes. II. A simple analytic prescription}",
    eprint = "2011.03014",
    archivePrefix = "arXiv",
    primaryClass = "astro-ph.CO",
    doi = "10.1103/PhysRevD.103.063538",
    journal = "Phys. Rev. D",
    volume = "103",
    number = "6",
    pages = "063538",
    year = "2021"
}

@article{Pi:2017gih,
    author = "Pi, Shi and Zhang, Ying-li and Huang, Qing-Guo and Sasaki, Misao",
    title = "{Scalaron from $R^2$-gravity as a heavy field}",
    eprint = "1712.09896",
    archivePrefix = "arXiv",
    primaryClass = "astro-ph.CO",
    reportNumber = "YITP-17-135",
    doi = "10.1088/1475-7516/2018/05/042",
    journal = "JCAP",
    volume = "05",
    pages = "042",
    year = "2018"
}

@article{Gangopadhyay:2026xqj,
    author = "Gangopadhyay, Mayukh R.",
    title = "{One Feature, Three Clocks: Phase-Locked Gravitational Waves, Primordial Black Holes, and Non-Gaussianity from Periodic Warm Inflation}",
    eprint = "2606.31430",
    archivePrefix = "arXiv",
    primaryClass = "astro-ph.CO",
    month = "6",
    year = "2026"
}

@article{Ning:2026jkk,
    author = "Ning, Zhuan and Cai, Rong-Gen and Wang, Shao-Jiang and Yoo, Chul-Moon",
    title = "{Long-term 3+1 simulations of primordial black hole formation during radiation domination}",
    eprint = "2608.13206",
    archivePrefix = "arXiv",
    primaryClass = "gr-qc",
    reportNumber = "NU-QG-25",
    month = "8",
    year = "2026"
}

@article{Clarke:2020bil,
    author = "Clarke, Thomas J. and Copeland, Edmund J. and Moss, Adam",
    title = "{Constraints on primordial gravitational waves from the Cosmic Microwave Background}",
    eprint = "2004.11396",
    archivePrefix = "arXiv",
    primaryClass = "astro-ph.CO",
    doi = "10.1088/1475-7516/2020/10/002",
    journal = "JCAP",
    volume = "10",
    pages = "002",
    year = "2020"
}

@article{Chandrasekhar:1931ih,
    author = "Chandrasekhar, Subrahmanyan",
    title = "{The maximum mass of ideal white dwarfs}",
    doi = "10.1086/143324",
    journal = "Astrophys. J.",
    volume = "74",
    pages = "81--82",
    year = "1931"
}

@article{Maiti:2025ijr,
    author = "Maiti, Subhasis and Maity, Debaprasad",
    title = "{The magnetic origin of primordial black holes: a viable dark matter scenario}",
    eprint = "2508.19217",
    archivePrefix = "arXiv",
    primaryClass = "astro-ph.CO",
    doi = "10.1088/1475-7516/2026/04/020",
    journal = "JCAP",
    volume = "04",
    pages = "020",
    year = "2026"
}

@article{Maiti:2026hsn,
    author = "Maiti, Subhasis",
    title = "{Gravitational waves from postinflationary magnetism: Direct and scalar-induced contributions}",
    eprint = "2605.24715",
    archivePrefix = "arXiv",
    primaryClass = "astro-ph.CO",
    doi = "10.1103/8jkm-5l84",
    journal = "Phys. Rev. D",
    volume = "114",
    number = "4",
    pages = "043530",
    year = "2026"
}

\end{document}